\documentclass[aps,onecolumn,prd,showpacs,showkeys,preprintnumbers,superscriptaddress,nobibnotes,notitlepage,floatfix,longbibliography,nofootinbib]{revtex4-2}
\usepackage[utf8]{inputenc}
\usepackage{amssymb}
\usepackage{amsmath}
\usepackage{graphicx}
\usepackage{xcolor}
\usepackage{multirow}
\usepackage{hyperref}
\usepackage{diagbox}

\def\tev{\, {\rm TeV}}
\def\gev{\, {\rm GeV}}

\def\met{\, {\rm MET}}

\newcommand{\gsim}{\lower.7ex\hbox{$\;\stackrel{\textstyle>}{\sim}\;$}}
\newcommand{\lsim}{\lower.7ex\hbox{$\;\stackrel{\textstyle<}{\sim}\;$}}
\newcommand{\fb}{\rm fb}
\newcommand{\ifb}{\rm fb^{-1}}

\newcommand\be{\begin{equation}}
\newcommand\ee{\end{equation}}
\newcommand\bea{\begin{eqnarray}}
\newcommand\eea{\end{eqnarray}}

\begin{document}
\preprint{MI-HET-810, HRI-RECAPP-2023-08, UH511-1330-2023, CETUP-2023-007}
\title{Machine Learning Techniques for Intermediate Mass Gap\\
Lepton Partner Searches at the Large Hadron Collider}

\author{Bhaskar Dutta}
\affiliation{Department of Physics and Astronomy,
Mitchell Institute for Fundamental Physics and Astronomy,\\
Texas A\&M University, College Station, TX  77843, USA}

\author{Tathagata Ghosh}
\affiliation{Harish-Chandra Research Institute,
A CI of Homi Bhabha National Institute,\\
Chhatnag Road, Jhusi, Prayagraj 211019, India}

\author{Alyssa Horne}
\affiliation{Department of Physics, Sam Houston State University, Huntsville, TX 77341, USA}
\affiliation{Department of Physics, Michigan Technological University, Houghton, MI 49931, USA}

\author{Jason Kumar}
\affiliation{Department of Physics and Astronomy, University of Hawai'i, Honolulu, HI 96822, USA}

\author{Sean Palmer}
\affiliation{Department of Physics, Sam Houston State University, Huntsville, TX 77341, USA}
\affiliation{Department of Physics, New Mexico State University, Las Cruces, NM 88003, USA}

\author{Pearl Sandick}
\affiliation{Department of Physics and Astronomy, University of Utah, Salt Lake City, UT  84112, USA}

\author{Marcus Snedeker}
\affiliation{Department of Physics, Sam Houston State University, Huntsville, TX 77341, USA}
\affiliation{Department of Physics, East Carolina University, Greenville, NC 27858, USA}

\author{Patrick Stengel}
\affiliation{Istituto Nazionale di Fisica Nucleare, Sezione di Ferrara, via Giuseppe Saragat 1, I-44122 Ferrara,
Italy}

\author{Joel W.~Walker}
\affiliation{Department of Physics, Sam Houston State University, Huntsville, TX 77341, USA}


\begin{abstract}
We consider machine learning techniques
associated with the application of a Boosted Decision Tree (BDT) to searches at the Large 
Hadron Collider (LHC) for pair-produced lepton partners which decay to leptons and invisible 
particles.  This scenario can arise in the Minimal Supersymmetric Standard Model (MSSM), but 
can be realized in many other extensions of the Standard Model (SM).  We focus on the case of intermediate mass 
splitting ($\sim 30 \gev$) between the dark matter (DM) and the scalar.  For these mass splittings, the LHC has made little improvement over LEP due to large electroweak backgrounds.
We find that the use of machine learning techniques can push the LHC well past
discovery sensitivity for a benchmark model with a lepton partner mass of $\sim 110 \gev$, 
for an integrated luminosity of 
$300~\fb^{-1}$, with a signal-to-background ratio of $\sim 0.3$.  
The LHC could exclude models with a lepton partner mass as large as $\sim 160 \gev$ with the 
same luminosity.  
The use of machine learning techniques in searches for scalar lepton partners at the LHC could thus definitively probe the parameter space of the MSSM in which scalar muon mediated interactions between SM muons and Majorana singlet DM can both deplete the relic density through dark matter annihilation and satisfy the recently measured anomalous magnetic moment of the muon.
We identify several machine learning 
techniques which can be useful in other LHC searches
involving large and complex backgrounds.
\end{abstract}

\maketitle

\section{Introduction}

A wide variety of scenarios for physics beyond the Standard Model (BSM) have been 
probed experimentally with the Large Hadron Collider (LHC).
However, as no conclusive evidence of BSM physics has been
found yet, focus has turned to scenarios and 
signatures which are more difficult to probe.
One particularly difficult scenario is the 
pair production of scalar lepton partners of SM fermions
($\tilde \ell^\pm)$, each of which decays 
to a lepton and an invisible particle 
($\tilde \ell \rightarrow \ell X$) with a 
$\sim30\gev$ mass splitting.  This 
scenario can arise in the minimal supersymmetric 
standard model (MSSM)~\cite{Fukushima:2014yia,Acuna:2021rbg}, as well as in other BSM models~\cite{Acuna:2021rbg,Acuna:2022ouv} often 
specifically motivated by the need for a viable dark matter candidate and 
by recent measurements of the  anomalous magnetic moment of the muon.
This scenario is difficult to probe at the LHC 
because there is a large SM background from the 
production of electroweak gauge bosons, decaying 
to leptons and neutrinos.  As a result, current 
LHC constraints~\cite{ATLAS:2022hbt,CMS:2018eqb} on this scenario show little 
improvement over LEP~\cite{LEP}.   Although a variety 
of new analysis strategies have been proposed, 
there is no clear-cut strategy for effectively 
separating 
signal from background in models with these moderately compressed particle spectra.  In this work, we 
investigate the possibility of using machine 
learning algorithms to more effectively pick out 
signal from background.  

The standard signature of scalar lepton partner pair production 
at the LHC is $\bar \ell \ell$ plus missing transverse energy ${/\!\!\!\!E}_{\rm T}$ (MET).  However, if the mass 
splitting is $\lesssim 60~\gev$, then the lepton and 
invisible particles are both soft, and it is difficult 
to see this signature above the electroweak background~\cite{ATLAS:2019lff}.  
One strategy used to evade the difficulty of soft particles 
is to search for events in which one or more hard jets are 
also emitted ($pp \rightarrow \tilde \ell^* \tilde \ell 
+~{\rm jets}$).  This jet provides a transverse kick to the 
$\tilde \ell^* \tilde \ell$ system, yielding harder 
leptons and larger MET.  Even so, distinguishing signal 
events from the large electroweak background requires 
additional techniques.  If the mass splitting is small 
($\lesssim 20~\gev$), then the electroweak background can 
be rejected because the neutrinos produced by $W$ decay 
typically carry larger MET than carried by the $X$ arising 
from $\tilde \ell$ decay~\cite{ATLAS:2019lng}.  The efficacy of this search 
strategy has been enhanced by recent experimental 
developments, resulting in lower energy thresholds for 
lepton identification~\cite{ATLAS:2020ofx}.
However, searches at the LHC are still challenging
for intermediate mass splittings in 
the range of $\sim 20-50~\gev$.

Theoretical work has shown that progress can be made for 
mass splittings in the $\sim 20-50~\gev$ range, using a 
variety of kinematic variables involving the energies of 
the leptons, jets, and MET, as well as their angular 
correlations with each other~\cite{Dutta:2017nqv}.  But the progress is 
incremental; even applying these strategies, a 
model with $M_{\tilde \ell} = 160~\gev$ and 
$M_{\tilde \ell} - M_X = 30~\gev$ 
would be well out of reach of LHC with $300~\ifb$ luminosity~\cite{Dutta:2017nqv}.   
Moreover, there is no clean set of simple cuts which one 
can apply to maximize sensitivity, based on an a priori 
principle.  Instead, one sequentially applies 
cuts to many kinematic variables, with each cut (and the 
order of cuts) being 
determined largely by trial and error. This is the type of setting in which one might expect 
machine learning algorithms to greatly improve search 
sensitivity, by determining the optimal set of cuts to 
impose on a large number of kinematic variables 
which can exhibit strong (nonlinear) correlations.
Additionally, one might hope that
it would be possible to work backwards 
after an optimal set of cuts is found
to determine the underlying principle(s)
responsible for the efficacy of those cuts.
We will see that both of these intuitions are true.  

In this work, we consider a benchmark point which is allowed by current constraints (cf.~discussion in Section~\ref{sct:snb})
where the scalar lepton is a partner to the left-handed muon, with 
$M_{\tilde \ell} = 110~\gev$, $M_{\tilde \ell} - M_{\chi} =30~\gev$.
We use a boosted decision tree (BDT)~\cite{BDT:2000,BDT:2001} to classify events as signal or background,
finding that a cut based on the BDT classifier can increase the LHC sensitivity
to $\gtrsim 5\sigma$ while maintaining a signal-to-background ratio of $\sim 0.3$ 
with $300~\ifb$ of data. 
We identify kinematic variables
that dominantly contribute to the signal sensitivity and plot the residual distribution as applicable.

Importantly, we find that direct application of a BDT algorithm to simulated data is
not sufficient.  Essentially, because some SM background processes have very 
large rates, application of a BDT algorithm to an uncurated data sample will 
succeed at removing the largest backgrounds, while leaving subleading backgrounds that, in turn, can limit sensitivity.  Instead, we adopt a strategy in which some initial precuts 
are imposed (in addition to the basic cuts which define the event topology) to reduce leading backgrounds with easily identified characteristics,
leaving the BDT free to focus subsequently on more challenging features of
the residual leading and subdominant backgrounds. 

In a similar vein, we find that generation of the training sample used by the BDT also requires special care,
because the most difficult backgrounds to remove lie in regions of phase space 
with relatively small cross section (though larger than the signal).  If these regions of phase 
space are not adequately sampled by simulation, then the BDT can become overly focused on the 
details of a few particular events in the training sample.  To avoid this problem, we simulate in kinematic tranches, significantly enhancing the fraction of computational time
devoted to kinematic regions having smaller cross sections but larger pass rates for standard cuts.

In previous studies, a variety of machine learning techniques have been proposed and implemented
for analyses of LHC data in which the BSM signal is particularly challenging to differentiate from the SM background.
Following an early proposal for the use of BDTs and neural networks (NNs) in searches for
fermionic partners of SM electroweak gauge bosons~\cite{Baldi:2014kfa}, 
the pair production of such fermionic partners
has been constrained in a BDT analysis of $(\bar \ell \ell +\met )$ final states at the LHC~\cite{ATLAS:2022hbt}.
For boosted topologies, both BDT and NN analyses have been shown to enhance the
sensitivity of LHC searches to dark matter models with extended Higgs sectors~\cite{Dey:2019lyr}.
Analyses of monojet plus missing energy signatures at LHC in several simplified dark matter
models with mediators of different spins has demonstrated that NNs are able to efficiently
determine if there is a BSM signal within a given dataset when trained on two-dimensional
histograms combining information from different kinematic variables~\cite{Khosa:2019kxd,Arganda:2021azw}. 

In addition to supervised learning applications, less supervised
techniques have been proposed for BSM searches at LHC. Weakly supervised techniques have been
shown to increase the sensitivity of monojet plus missing energy searches to strongly
interacting dark matter models with anomalous jet dynamics~\cite{Finke:2022lsu} and
self-supervised contrastive learning has been proposed for anomaly detection searches in dijet events~\cite{Dillon:2022tmm}.
Complementing these largely model-agnostic techniques, adversarial NNs have also been suggested as an unsupervised approach to improve the invariant mass reconstruction of BSM gauge bosons decaying to leptons and neutrinos~\cite{MB:2023edk}.

In this work, we favor an application of supervised learning strategies
using a BDT due to the relatively straightforward manner in which
the algorithm classifies signal and background events,
facilitating a clear interpretation of results.
We will focus not only on the search sensitivity which is achieved by application of a BDT,
but also on what the BDT teaches us regarding the underlying search strategy,
and on improved techniques for applying machine learning to general LHC searches with large and complex backgrounds.

The search strategy we propose is relevant for a variety of phenomenologically motivated extensions of the SM.
For example, charged mediator models with a Majorana SM-singlet dark matter candidate and
scalar muon partners mediating interactions with the SM can both satisfy the dark
matter relic density and account for the  anomalous magnetic moment of the muon~\cite{Ghosh:2022zef}.
With the coupling of the dark matter to the muon and its scalar partner fixed by the SM
hypercharge coupling, as in the MSSM, the relic density in such models can be
depleted by scalar mediated dark matter annihilation for lepton partner masses
$M_{\tilde \ell} \lsim 150~\gev$~\cite{Fukushima:2014yia} and by co-annihilation processes
involving the scalars for $M_{\tilde \ell} \lsim 1~\tev$~\cite{Acuna:2021rbg}.
For simplified models with dark matter-scalar-lepton couplings larger than in the MSSM,
both production mechanisms can yield the observed dark matter relic density for scalar
masses $M_{\tilde \ell} \gsim 1 ~\tev$~\cite{Acuna:2021rbg,Acuna:2022ouv}.
Charged mediator models with the most massive scalar muon partners
mentioned above are virtually impossible to detect at LHC due to the falling
production cross sections at higher scalar masses. However, there are large
regions of unconstrained parameter space which are challenging but feasible to
probe for LHC searches with the improved sensitivity of analyses using machine learning techniques such as a BDT. 

The plan of this paper is as follows.  In Section~\ref{sct:motivation}, we describe our approach and motivation for using a BDT.
We outline the scalar muon partner signal and associated backgrounds, along with a summary of previously implemented and proposed analyses using cut-and-count strategies, in Section~\ref{sct:snb}. The details of the event simulation, selected observables, BDT training, and results of our analysis are described in Section~\ref{sct:bdt}. We discuss the conventional wisdom which can be gained from our study and applied to others in Section~\ref{sct:conventionalwisdom}, before summarizing our findings and briefly discussing future work in Section~\ref{sct:conclusion}.

\section{Background and Motivation for Using a BDT}
\label{sct:motivation}

By its nature, known physics is necessarily more prevalent or more readily
visible to conventional experimental techniques than unknown physics,
while unknown physics and the opportunity to extend our understanding
of fundamental particles and interactions are of greater basic interest.
Accordingly, the problem of how one efficiently suppresses known backgrounds
in order to enhance the prospective visibility of new processes is of great interest.
In a prior study~\cite{Dutta:2017nqv}, we attempted to systematically separate the signal associated with the production of lepton-partners in intermediate mass gap scenarios from competing SM backgrounds via an iterative process involving the
plotting of distributions in relevant kinematic
observables and the manual application of event selection cuts.
This approach proved effective, although there are a number of
ways in which it is potentially sub-optimal.  

Firstly, it was observed that this process is extremely sensitive
to the order in which the cuts are applied, since a variety of
useful secondary and tertiary event selections remain obscured
until a majority of foreground debris is removed by primary selections.
A corollary of this statement is that the possible sequencing of cuts
explodes combinatorically, and it is very difficult to be certain that
a given sequence of cuts is in any sense optimal.
Likewise, there is a great danger that early event selections can be
applied too aggressively, precluding more surgically targeted cuts
down the line.  This is part of a larger problem, that any by-hand
procedure is intrinsically ad hoc and one-off,
while simultaneously remaining extremely labor-intensive.
Finally, there is a risk of biasing the analysis in the limit
of low statistics, which always ultimately applies as selections
become increasingly strict, since cuts are typically engineered
and validated with respect to the same collection of events.

Machine learning offers substantial promise for alleviating
negative outcomes associated with each of these objections.
Specifically, it can much more thoroughly and efficiently
scan the space of available discriminants, while delivering
results that are more stable and more reproducible, using a 
generalizable approach that has increased cross-applicability,
and which requires less investment of human capital
into repetitive and automatable tasks.  Additionally,
leading machine learning algorithms offer built-in protection against
over fitting.  For example, the learning rate can be turned down
such that the likelihood of over-applying an early selection and
losing sensitivity to future gains is reduced.  Likewise,
the training and testing samples can be readily isolated,
reducing the risk of learning random features of the presented
sample that are not replicated in the larger ensemble.
Also, the separation between training and testing can typically
be ``folded'' in multiple ways, allowing for cross-validation
via replication of the analysis, which helps to quantify the
stability of results against statistical fluctuations.
Much greater complexity and refinement of the
discriminant is available in a machine learning context relative
to manual approaches.  Additionally, the assignment of a
continuous event-by-event classification likelihood represents
a much richer category of information than the discretization
implicit to a more basic cut-and-count approach.

However, a pressing concern associated with machine learning
is that one often sacrifices the ability to investigate \emph{what}
was learned, and to develop intuition regarding the nature of the
signal and background separation. This is a key reason that we
favor the use of supervised learning with a BDT rather deep learning approaches utilizing more complex algorithms such as NNs.\footnote{While deep NNs are typically considered to be less inherently interpretable than a BDT, there has been significant recent work on both algorithm-agnostic and NN-specific methods to evaluate the explainability of deep learning algorithms (for a recent overview, see Ref.~\cite{XAIsurvey:2020}).}
In particular, the input to a BDT is a simple list
of numerical data associated with each event, together with an event
weight and a known classification in the case of the training sample.
The BDT goes through a process that is similar to the
design of a manual event selection profile in many regards, but it is
strictly systematic, mathematically rigorous, and reproducible.
At each stage of training, the optimal discriminant and splitting value is
ascertained according to a well-defined loss function; subsequently,
each branch of this ``tree'' can elect its own best choice for the
next splitting, and so on.  Likewise, the classification score assigned
to each terminal leaf of the tree is selected by extremization of
the same functional, in the effort to optimally reconcile feature-based
predictions with corresponding truth labels.
The capacity to differentially select supplementary discriminants for
previously separated populations adds a level of flexibility and
refinement that is quite challenging to implement by hand.

A number of such trees are successively generated
in the ``boosting'' process, wherein events are reweighted to emphasize
the correction of prior missorts in subsequent refinements of the training.
By combining a long sequence of relatively shallow trees, a strong
discriminant is built from the conjunction of many ``weak learners.''
Adjustable ``regularization'' factors built into the training objective may be
tuned to veto branchings that would add complexity without meaningfully
reducing the classification loss, and to slow the rate of learning
to limit the danger of over fitting.
Training may also be halted prior to completion of the specified
maximal tree count if real-time validation against the testing sample
indicates an onset of diminishing returns via a plateau in the loss function. 
Collectively, these measures help to mitigate the so-called
``bias-variance'' problem, i.e. the problem of over training
on features that are not widely generalizable.
Ultimately, the final classification score assigned to a given event
represents the sum over its leaf scores for all trees.  This
unbounded value $y$ is typically mapped onto the bounded range $p(y)\in \{0,1\}$
via application of the sigmoid ``logistic'' function, or similar.
Crucially, the process by which the classification scores are assigned
is entirely tractable, and one may output the full set of constructed
decision trees if so desired, including the splitting features and values,
as well as the selected leaf scores.  In addition, a summary report of the
``importance'' of each feature to the training is readily available, and it is possible
to iteratively reduce the dimensionality of the provided feature space using this information,
selectively eliminating redundant (highly correlated) and indiscriminant features.

Rareness of the targeted new physics processes, and likewise, of the most competing SM backgrounds
(in the tail regions of the background distributions), implies that it is generally computationally impossible
to work with simulated events in their naturally occurring proportions.  The
solution is to keep track of per-event weights, i.e. partial cross-sections
linked to each event that represent the
extent of the final-state phase space for which they are a working proxy.
Even within a category of final state particles, e.g. top quark pair production
or dibosons plus jets, events representing the higher-energy hard scatterings that
are generally of greater interest after an application of selection cuts
are generally power-suppressed relative to their softer low-energy counterparts.
Accordingly, it becomes very computationally inefficient to simulate enormous 
quantities of events that are dominated by softer scatterings that will
mostly be discarded in order to emphasize harder events in the distribution tails.
This can likewise be handled by separating each simulation category into
disjoint tranches that are binned in some relevant process scale such
as the scalar sum over transverse momentum $p_{\rm T}$.
A significant quantity of rarer events can be generated at lower cost in this way, 
allowing the analysis to maintain a more uniform level of statistical
representation across the phase space.  The rare events then simply carry
smaller per-event weights, which can be fed into the machine learning in order
to direct its attention appropriately toward the most proportionally relevant features at a given stage of the training.

Although more sophisticated types of deep learning such as neural networks (NNs) can potentially outperform BDTs in the classification of signal and background events, previous studies suggest the performance of NNs and BDTs are comparable. In a study of boosted event topologies relevant for dark matter models with extended Higgs sectors~\cite{Dey:2019lyr}, BDT and NN implementations yield similar increases in sensitivity relative to a cut-based analysis. A comparison between deep NNs and BDTs in Ref.~\cite{Baldi:2014kfa} demonstrates only marginally better performance by NNs in an analysis of scalar lepton pair production, without the boost from a hard jet. A similar comparison in searches for fermionic partners of electroweak gauge bosons shows a more significant improvement in performance by deep NNs compared to BDTs, which is attributed to the lack of high-level kinematic variables, whose larger discriminating power is necessary for an optimal BDT analysis. A dedicated study of NNs implemented for the pair production of scalar leptons in boosted event topologies at LHC is thus an interesting possibility for future work.

\section{Characterizing Signal and Leading Backgrounds}
\label{sct:snb}
For our analysis, we focus on a model in which the scalar lepton $\tilde \ell$ is the partner of the left-handed muon,
with $M_{\tilde \ell} = 110~\gev$ and $M_{\tilde \ell}-M_X=30~\gev$ ($\ell = \mu$).  
We assume $X$ is a stable SM-singlet Majorana fermion.  
This scenario is thus realized in the MSSM, where $\tilde \ell$ is a left-handed 
smuon, $X$ is a bino-like lightest supersymmetric particle (LSP), and other SUSY particles are relatively 
heavy. However, this scenario can also be realized in a variety of other phenomenologically well-motivated BSM models.
The event topology we are interested in is $pp \rightarrow \bar \ell \ell +\met + {\rm jets}$.  
This topology can 
be produced by signal events in which proton collisions result in $\tilde \ell^* \tilde \ell$ pair production, 
with each new scalar decaying via $\tilde \ell \rightarrow \ell X$, with one or more jets.
We simulate the signal process using the {\tt MSSM\_SLHA2}~\cite{Allanach:2008qq} model distributed with {\sc MadGraph}~\cite{Alwall:2014hca}.
Model parameters are generated consistently using the {\sc SUSY-HIT}~\cite{Djouadi:2006bz} package,
with benchmark smuon and neutralino masses taking the values indicated previously.
In addition to initial-state radiation, jets can also be produced as part of the hard scattering, and we simulate the signal process inclusively, with 0, 1, or 2 final-state jets (with matching).  Two example Feynman diagrams (from {\sc MadGraph}) are shown in FIG.~\ref{fig:feynman}.

\begin{figure}[htpb]
\centering
\includegraphics[width=0.4\linewidth]{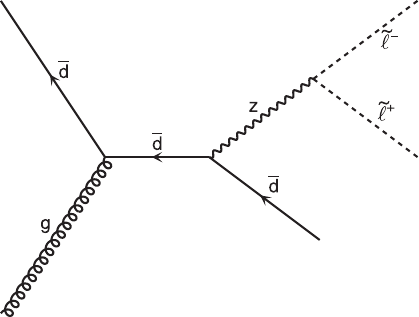}
\quad \quad
\includegraphics[width=0.4\linewidth]{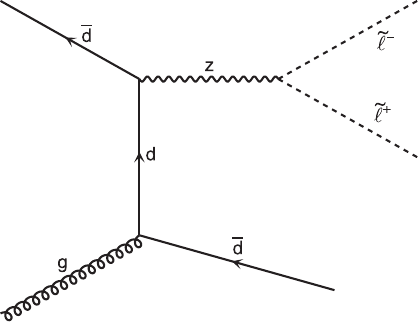} 
\caption{Two example Feynman diagrams representing signal production with an associated jet.}
\label{fig:feynman}
\end{figure}

This region of parameter space is still allowed by analyses of $139~\fb^{-1}$ of ATLAS data~\cite{ATLAS:2022hbt,ATLAS:2019lng}, 
which did not utilize machine learning techniques. Indeed, for a mass splitting of $30 \gev$, the tightest constraint is still from 
LEP~\cite{LEP}, which rules out the region of parameter space with $M_{\tilde \ell} \lesssim 97 \gev$ for right-handed smuons.\footnote{LEP searches did not attempt to constrain left-handed smuons due to the relatively heavy masses expected in concrete realizations of the MSSM. Since LEP had sufficient luminosity for the right-handed smuon searches to be kinematically limited by the center-of-mass energy of the beamline, the constraints on left-handed smuon masses are expected to be similar due to the larger cross sections relative to right-handed smuon production.}
A scenario with similar phenomenology is explored in Ref.~\cite{Ashanujjaman:2023tlj}.

The leading SM backgrounds to this signal arise from processes in which the charged leptons are produced 
from the decay of on-shell weak gauge bosons, with missing energy arising either from neutrinos 
or jet mismeasurement.
The main such processes are
\begin{itemize}
\item{$pp \rightarrow Zj$, with $Z \rightarrow \bar \tau \tau, \bar \ell \ell$.  If $Z$ decays to $\bar \tau \tau$, then 
the $\tau$s decay to $\ell$, $\bar \ell$ and neutrinos, while if $Z$ decays to $\bar \ell 
\ell$, then MET arises from mismeasurement of the jet energy; }
\item{$pp \rightarrow \bar t t$, with $t \rightarrow bW$, and the $W$s decaying to $\ell$, $\bar \ell$ and neutrinos;}
\item{$pp \rightarrow ZZ, W^+ W^-$, with the massive gauge boson decays producing $\ell$, $\bar \ell$ and neutrinos;}
\item{$pp \rightarrow \bar \tau \tau j$, with the $\tau$s decaying to $\ell$, $\bar \ell$ and neutrinos;}
\item{$pp \rightarrow WZ$, with the massive gauge boson decays producing a neutrino and three charged leptons, with one 
charged lepton missed.}
\end{itemize}
Of these background processes, the rate for 
$pp \rightarrow Zj$ dominates by roughly two orders of magnitude over the nearest competitor.  But the 
rates for all of these processes are much larger than for $pp \rightarrow \tilde \ell^* \tilde \ell$.  Thus, to obtain significant 
improvements in sensitivity and signal-to-background ratio, 
it is necessary to strongly reject the dominant background, while still retaining strong rejection of 
the subleading backgrounds. 

Because $pp \rightarrow Z j$ events largely yield a $Z$ near rest, the neutrinos produced from the products 
of the $Z \rightarrow \bar \tau \tau$ decay process tend to have low energy.  As a result, a requirement of 
minimum missing energy is successful in reducing this background.  The requirement of a minimum 
missing energy also reduces the background arising from $pp \rightarrow Zj$, with $Z \rightarrow \bar \ell \ell$, 
since the likelihood that MET arises from jet mismeasurement falls with increasing MET.
$pp \rightarrow Zj$ events can also be distinguished by the kinematic variables built from 
the lepton and jet momenta, including the dilepton invariant mass ($M_{\ell \ell}$) and the ditau invariant  mass ($M_{\tau \tau}$), which will be described further in the next section.
The utility of the $M_{\tau \tau}$ variable arises from the fact that, in the 
$Z \rightarrow \bar \tau \tau$ process, the $\tau$s are boosted.  As a result, the neutrinos produced by $\tau$-decay are 
largely collinear with the charged leptons, allowing one to reconstruct the mass of the parent particle 
using transverse momentum conservation.  The difficulty lies not so much in removing the $Z$-related background, but 
in doing so without compromising one's ability to distinguish signal from the remaining backgrounds.  
In~\cite{Dutta:2017nqv}, the cut-and-count strategy for reducing $Z$-related backgrounds focused at the level of primary selections on: 
\begin{itemize}
\item{Rejecting events with $M_{\ell \ell}$ within  $M_Z \pm 10 \gev$,  }
\item{Rejecting events with $M_{\tau \tau} < 125 \gev$, and }
\item{Rejecting events with $\met < 125 \gev$.}
\end{itemize}
Additional kinematic variables can be used to discriminate  between signal and $W$- and top-related backgrounds, 
based on the energy and angular distribution of the decay processes.  In~\cite{Dutta:2017nqv}, a total of seven secondary and tertiary cuts were applied in sequence to define a signal region.  We will see that a BDT analysis can provide substantial improvement relative to that approach.

\section{Simulation and BDT Analysis} \label{sct:bdt}

\subsection{Event Simulation}

Monte Carlo training samples were generated for the
$\sqrt{s} = 13$~TeV LHC\footnote{Additional data collected by ATLAS and CMS up to and beyond $\mathcal{L} = 300 \, \ifb$ will be at higher center-of-mass energy, $\sqrt{s} \lsim 14$~TeV. We consider $\sqrt{s} = 13$~TeV for ease of comparison to previous analyses and note that the production cross-sections for all relevant signal and background processes vary by $\lsim 10\%$ between $\sqrt{s} = 13$~TeV and $\sqrt{s} = 14$~TeV. All processes are calculated at tree-level since, as discussed in Ref.~\cite{Dutta:2017nqv}, K-factors which account for higher order corrections have little impact on the projected sensitivity.} using {\sc MadGraph}/{\sc MadEvent}~\cite{Alwall:2014hca},
with showering and hadronization
in {\sc Pythia8}~\cite{Sjostrand:2014zea},
and detector simulation in {\sc Delphes}~\cite{deFavereau:2013fsa}.
We consider events in which one finds a $\mu^+ \mu^-$ pair, exactly one hard central jet ($P_{\rm T} \ge 30$~GeV, $\vert\eta\vert <2.5$), zero $b$-tagged jets, MET ($\ge 30$~GeV), and no hadronic $\tau$ decays.\footnote{We assume, following~\cite{ATLAS:2022hbt,ATLAS:2014cva}, 
that the effects of pile-up can be reduced 
by requiring that, for jets with $p_T < 60~\gev$, a sufficient fraction of the tracks associated with the 
jet point back to the primary vertex, as identified by the jet vertex tagger.}
We refer to these defining event selections as the ``topology cuts'', and they
correspond to the primary event selections described Ref.~\cite{Dutta:2017nqv}.
They are distinguished from the further ``precuts'' to be itemized subsequently,
which are also applied manually, and prior to the main BDT analysis.

We consider six major background processes: $\bar \mu \mu jjj$, $\bar \tau \tau jjj$, $\bar t t jj$, $WWjj$, $ZZjj$, $WZjj$.
Events are simulated inclusively, with up to two or three additional jets, depending on the process (as summarized in Table~\ref{tab:xsec}).  Specifically, we combine processes with zero up to the specified maximal number of hard isolated jets $j$ at the Feynman diagram level (in {\sc MadGraph}), and perform jet matching (including the simulation of initial state radiation and related effects) in conjunction with {\sc Pythia8}.~\footnote{%
For example, the $\bar \mu \mu jjj$ process card includes the instructions {\tt generate p p > l+ l-},
{\tt add process p p > l+ l- j}, {\tt add process p p > l+ l- j j}, and {\tt add process p p > l+ l- j j j}.
These processes are simulated by {\sc MadGraph} and are reliable for hard, well-separated partonic constituents.
Showering is performed by {\sc Pythia8},
and is reliable in the complementary limit of soft and/or collinear radiation
(where it may generate any number of partonic states).
The matching procedure allows each approach to handle its realm of specialization 
while partitioning the phase space to avoid double counting.}
Note that $Z$+jets backgrounds are already included in the $\bar \mu \mu jjj$ and  $\bar \tau \tau jjj$ 
classes, which also include lepton production through off-shell $Z^*$ and photon $\gamma^*$ mediators.
We do not directly simulate the production of $tW$,
whose final state is quite similar to that of $t\bar{t}$.
At production level, the inclusive $t\bar{t}+$~jets~\cite{CMS:2017xrt} background dominates
by more than an order of magnitude over that of $tW+$~jets~\cite{CMS:2018amb},
although that rate is offset to some extent by an extra opportunity for a $b$-jet veto.
One significant challenge to the inclusive simulation of $tW$ backgrounds is
that they are quite difficult to disentangle from double counting with $t\bar{t}$.
The cross sections for signal and background events with the stipulated topology
are given in the middle column of Table~\ref{tab:xsec}.  Note that cross sections
for the individual background processes vary widely, with all being larger than the signal.

\begin{table}[ht]
\centering
\begin{tabular}{|c|c|c|}
\hline
process & $\sigma$ (fb) (topology cuts) & $\sigma$ (fb) (precuts) \\
\hline
$\tilde \ell^* \tilde \ell$ & 12.3 & 1.38 \\
\hline
$\bar \mu \mu jjj$ & 12500 & 3.19 \\
$\bar \tau \tau jjj$ & 596 & 14.6 \\
$\bar t t jj$ & 66.8 & 5.49 \\
$ZZ jj$ & 26.6 & 0.234 \\
$ZW jj$ & 46.9 & 0.355 \\
$WW jj$ & 72.6 & 5.44 \\
\hline
\end{tabular}
\caption{Table of residual cross sections for the listed processes to have the correct topology 
(middle column) and to pass the precuts (right column).  The top row is the signal process, while 
the remaining rows are the leading backgrounds.}
\label{tab:xsec}
\end{table}

We tranche the simulation of each final state into non-overlapping bins of
phase space in order to ensure statistically reliable population of the kinematic tails.
For direct dilepton production (including $\bar \tau \tau jjj$),
we break the simulation runs up according to the generator-level
$P_{\rm T}$ of the leading lepton.
For the case of $\bar t t jj$, we tranche on $P_{\rm T}$ of the top quark.
To accomplish this, it is necessary to asymmetrically decay the $\bar t$ directly in {\sc MadGraph}, since generator-level cuts would otherwise be applied to both the particle and antiparticle.~\footnote{%
The associated process card instructions are {\tt generate p p > t t~, t~ > w- j},
{\tt add process p p > t t~ j, t~ > w- j}, and {\tt add process p p > t t~ j j, t~ > w- j}.}
The $t$ is decayed subsequently in {\sc Pythia}, as usual.
For $ZW jj$, we tranche on $P_{\rm T}$ of the $Z$-boson.
The same tranching is applied to $ZZ jj$ production, after decaying one of the two $Z$-bosons leptonically.
For $WW jj$, we require at least one of the $W$-bosons to decay leptonically, and again tranche on the leading (or only) leptonic $P_{\rm T}$.
The signal model is forced to decay to a final state including
a di-muon pair and tranched on $P_{\rm T}$ of the leading muon.
Several bins of increasing width were simulated for each of these final state discriminants, in order to represent soft, intermediate, and hard multi-TeV scattering processes.
Specifically, the numerical boundaries selected for this study
were 50, 100, 150, 200, 300, 400, 500, 750, 1000, 1500, 2000, 3000, and 4000~GeV.
In this manner, events were generated with a smooth distribution of per-event weights spanning around 6-8 orders of magnitude.
We verified explicitly that the associated production cross-sections summed over tranches are consistent with those obtained by a single joint simulation.
In total, after jet matching but prior to the application of topology cuts, more than 300 million candidate background events were generated, along with over five million events for the signal benchmark.
After precuts, the number of surviving event samples passed to the BDT for further analysis is reduced to around a half million signal events and around 200,000 background events.
The corresponding cross-sections are indicated in the last column of Table~\ref{tab:xsec}.

\subsection{Kinematic Variables}
We provide the BDT with 27 variables that are consistent with the final state topology of an opposite-sign, same-flavor dilepton, a hard (non-$b$) monojet, and MET.
This set includes a variety of sophisticated ``high level'' observables which are known to be effective
against various components of the SM background, as well as a number of more basic kinematic inputs
referencing the dilepton pair ($\ell_1,\ell_2$), the hard jet $j$,
and/or the MET.
The most difficult task of the BDT is to distinguish the decay process $\tilde \ell \rightarrow \ell X$ from 
the background processes which involve $W$-boson decay $W \rightarrow \ell \nu$.
We thus include several discriminants that are specifically constructed to emphasize
differences in the mass or spin of the parent particle or invisible particle.  
The set of computed observables is summarized in Table~\ref{tab:variables}. 
For a more detailed discussion on why some of these variables are relevant to unearth the underlying physics of the compressed spectra topology, we refer the reader to 
Ref.~\cite{Dutta:2017nqv}.
We do not make any effort to remove degeneracies in the feature set
since the BDT is intrinsically resilient to such correlations.
Event analysis, including application of selection cuts and
the computation of observables, is performed
with the {\sc AEACuS}~\cite{aeacus,Walker:2022iwa} package.

\begin{table}[ht]
\centering
\bgroup
\def\arraystretch{1.5}
\begin{tabular}{|c|c|}
\hline
\multicolumn{2}{|c|}{\quad Invariant Masses \quad} \\ 
\hline
$M_{\ell\ell}$ & dilepton system mass \\
$M_{j}$ & jet mass \\
\hline
\multicolumn{2}{|c|}{\quad Mass-Like Constructions \quad} \\ 
\hline
$M_{\rm T2}^0$ & ~massless invisible hypothesis~ \\
$M_{\rm T2}^{100}$ & ~massive invisible hypothesis~ \\
$M_{\tau\tau}$ & ~ditau mass~ \\
\hline
\multicolumn{2}{|c|}{\quad ~Transverse Scales~ \quad} \\ 
\hline
${/\!\!\!\!E}_{\rm T}$ & missing transverse energy \\
$H_{\rm T}$ & ~scalar sum of hadronic $P_{\rm T}$~ \\
$M_{\rm eff}$ & sum of ${/\!\!\!\!E}_{\rm T}$ and $H_{\rm T}$ \\
\hline
\multicolumn{2}{|c|}{\quad Transverse Momentum Values \quad} \\ 
\hline
${P}_{\rm T}^{\ell_1}$ & leading lepton \\
${P}_{\rm T}^{\ell_2}$ & subleading lepton \\
${P}_{\rm T}^{j}$ & jet \\
\hline
\multicolumn{2}{|c|}{\quad Scale Ratios \quad} \\ 
\hline
\quad $(M_{\rm T2}^{100}-100) \div M_{\rm T2}^0$ \quad & ratio of excess mass \\
${P}_{\rm T}^{\ell_1}\div {E\!\!\!\!/}_{\rm T}$ & leading lepton to MET \\
${P}_{\rm T}^{\ell_2}\div {E\!\!\!\!/}_{\rm T}$ & subleading lepton to MET \\
${P}_{\rm T}^{j}\div {E\!\!\!\!/}_{\rm T}$ & jet to MET \\
\hline
\multicolumn{2}{|c|}{\quad Azimuthal Angular Separations \quad} \\ 
\hline
$\Delta \phi_{\ell_1 {/\!\!\!\!E}_{\rm T}}$ & leading lepton to MET \\
$\Delta \phi_{\ell_2 {/\!\!\!\!E}_{\rm T}}$ & ~subleading lepton to MET~ \\
$\Delta \phi_{j\,{/\!\!\!\!E}_{\rm T}}$ & jet to MET \\
$\Delta \phi_{\ell_1 \ell_2}$ & dilepton system \\
$\Delta \phi_{\ell_1 j}$ & leading lepton to jet\\
$\Delta \phi_{\ell_2 j}$ & subleading lepton to jet\\
\hline
\multicolumn{2}{|c|}{\quad Rapidity Separations \quad} \\ 
\hline
$\cos \theta^*$ & dilepton system \\ 
$\tanh \vert \Delta \eta_{\ell_1 j} \vert$ & leading lepton to jet\\
$\tanh \vert \Delta \eta_{\ell_2 j} \vert$ & subleading lepton to jet\\
\hline
\multicolumn{2}{|c|}{\quad Rapidity Values \quad} \\ 
\hline
$\eta_{\ell_1}$ & leading lepton \\
$\eta_{\ell_2}$ & subleading lepton \\
$\eta_{j}$ & jet \\
\hline
\end{tabular}
\egroup
\caption{Observables delivered to the BDT.}
\label{tab:variables}
\end{table}

Several of the itemized variables require further description.
In particular, $\cos \theta^*_{\ell_1 \ell_2}$~\cite{Barr:2005,Matchev:2012} 
is equal to the cosine of the polar scattering angle in the frame
where the pseudorapidities of the leptons are equal and opposite.
It is designed to reflect the fact that the angular distribution of intermediary
particles with respect to the beam axis in the parton center-of-mass frame
is determined by their spin, and that the lepton angular distribution should reflect this heritage.
Much of the practical utility of the $\cos \theta^*_{\ell_1 \ell_2}$ variable
hinges upon its resiliency against longitudinal boosts of the partonic system.
This feature is apparent in the definition $\cos \theta^*_{\ell_1 \ell_2} \equiv \tanh{(\Delta \eta_{\ell_1 \ell_2}/2)}$,
where $\Delta \eta_{\ell_1 \ell_2}$ is the pseudorapidity difference between the two leptons.
The $W$-boson associated backgrounds have a distribution in this variable that is almost flat up to a value around 0.8,
whereas distributions for the scalar-mediated signal models more sharply peak at zero, suggesting
a clear region of preference, as discussed in Ref.~\cite{Dutta:2017nqv}.
We construct analogous variables for other differences in pseudorapidity, as applicable.

The $M_{\rm T2}$~\cite{Lester:1999tx,Cheng:2008hk,Barr:2009jv} variable, which was used similarly in Ref.~\cite{Han:2014aea},
corresponds to the minimal mass of a pair-produced parent which
could consistently decay into the visible dilepton system and the observed
missing transverse momentum vector sum
under a specific hypothesis for the mass of the invisible species.
We compute two versions of this variable, $M_{\rm T2}^{0}$ and $M_{\rm T2}^{100}$,
corresponding respectively to a massless hypothesis and a $100$~GeV hypothesis.
The former is consistent with the MET arising from neutrinos, and the latter
is consistent with a dark matter candidate in the neighborhood of our benchmark model. 

Finally, the ditau mass variable $M_{\tau\tau}$~\cite{Ellis:1987xu} attempts to
reconstruct the invariant mass of a $\tau\tau$ pair that has decayed leptonically
under the hypothesis that the MET is associated with two pair of neutrinos
that are emitted collinearly with each of the observed leptons.
Momentum conservation in the transverse plane is then sufficient to reconstruct the energy of each neutrino pair, which in turn
determines the momentum of each hypothetical $\tau$ and allows one to reconstruct the invariant mass of the
pair.  This quantity may be expressed in closed form~\cite{Dutta:2017nqv} as
\be
M_{\tau \tau}^2 \equiv -M_{\ell_1 \ell_2}^2 \frac{(P_T^{\ell_1} \times P_T^j)\cdot (P_T^{\ell_2} \times P_T^j)}
{|P_T^{\ell_1} \times P_T^{\ell_2}|^2} \,,
\label{eq:mtautau}
\ee
where $M_{\ell_1 \ell_2}^2$ is the invariant squared mass of the visible lepton system.
One may confirm that $M_{\tau \tau}^2 = M_Z^2$ if the neutrinos are collinear with the charged leptons produced 
by $\tau$-decay, with the $\tau$s arising from the process $Zj \rightarrow \bar \tau \tau j$ 
(see Appendix~\ref{sec:Appendix_mtautau} for details).
Note that $M_{\tau \tau}^2 > 0$ if either $-P_T^{j}$ or $P_T^{j}$ lies between
$P_T^{\ell_1}$ and $P_T^{\ell_2}$,
and is negative if neither do.  This makes the ditau mass a good kinematic variable for more generally distinguishing event topology,
in addition to rejecting events that  involve the process $Z \rightarrow \bar \tau \tau \rightarrow \bar \ell \ell +4\nu$.
We define $M_{\tau \tau} \equiv {\rm sign}\,[M_{\tau \tau}^2]\times\!\sqrt{\vert M_{\tau \tau}^2\vert}$.

\subsection{Training} \label{sct:bdttraining}  

Since it is challenging to generate enough 
simulation data to represent LHC data in natural proportions,
simulated events are weighted based on the cross section of that background class as described previously.
Imbalances in the proportional representation of applicable backgrounds
lead to another difficulty in a BDT analysis: training samples are often dominated 
by backgrounds which are easily rejected by intuitively straightforward cuts, such as a dilepton mass cut around 
the $Z$-mass.  The BDT may thus learn very well how to reject a large set of background events for which a BDT was 
not really needed, since a simple cut would have done just as well.  On the other hand, the BDT may fail to 
adequately learn how to discriminate more difficult subleading backgrounds, which may likewise
be undersampled. Even if undersampling is mitigated by procedures similar to those described in the prior
subsection, training may be dominated by a small subset of events carrying large per-event weights. 

To address this issue, we impose a series of cuts, in addition to those which define the event topology, to cull data before the curated data is analyzed by the 
BDT.
These event selections are applied to the training data set and also the test data set, uniformly.
The goal of these initial precuts is to ensure that the surviving background event classes have roughly similar size, with a magnitude 
which is no more than $\sim {\cal O}(10-100)$ larger than the signal.  Essentially, these cuts have done the 
``easy work," allowing the BDT to focus on the hard work.  Moreover, these cuts provide an initial reduction 
in background based on principles which are known ab initio, allowing us to determine which kinematic variables 
the BDT uses to remove the more difficult backgrounds.

The precuts we use are:\footnote{While several analyses of LHC data (e.g.~\cite{ATLAS:2022hbt,CMS:2018amb,ATLAS:2021kqb}) have utilized a similar window cut on the invariant mass of the dilepton system around the mass of the $Z$-boson prior to implementing a BDT analysis, a more comprehensive investigation of precuts in such analyses has not been explored in previous studies.}
\begin{itemize}
\item{{\it Veto events with $M_{\mu \mu} \in 91 \pm 10\gev$}.  This cut dramatically reduces the 
$WZj$, $ZZj$ and $\mu^+ \mu^- j$ backgrounds 
by removing events in which the $\mu^+ \mu^-$ pair arise from $Z$-decay.}
\item{{\it Require ${\rm MET} > 110~\gev$}.  This cut reduces backgrounds, such as $\mu^+ \mu^- j$, in which 
MET arises from jet mismeasurement.} In addition, this cut also serves the dual purpose of acting as a trigger~\cite{ATLAS:2018trigger}.
\item{{\it Require $\cos \theta*_{\mu \mu} < 0.5$}.  This cut is effective in reducing the 
$WWj$ background, because it preferentially selects events in which the parent of the muon is 
spin-0, rather than spin 1.}
\end{itemize}
The cross sections for signal and background events which pass these precuts are given in the right column of Table~\ref{tab:xsec}.
After these precuts, the data is better suited for delivery to the BDT
for training on the discrimination of signal from background,
and better outcomes are achieved.
In particular, the maximal signal-to-background ratio is approximately doubled using this approach,  and the window for optimization of the classification score is significantly broadened,
implying improved stability of the result. 

Our BDT analysis and the generation of associated graphics
are done with the {\sc MInOS}~\cite{aeacus,Walker:2022iwa} package,
which calls {\sc XGBoost}~\cite{Chen:2016:XST:2939672.2939785} and 
{\sc MatPlotLib}~\cite{Hunter:2007} on the back end.
We train the BDT on 2/3 of the simulated event sample, reserving 1/3 for validation of outcomes
on a statistically independent sample.  We rotate the portion of data reserved for
validation in order to characterize statistical variations between each such data ``fold''.
We configure the BDT to build up to 50 trees with a maximal depth of 5 levels per tree.
Early stopping is enabled to halt training if the loss function is not improved 
over the course of the five most recent training epochs, i.e. by the generation
of the five most recent trees.

As described in Ref.~\cite{Chen:2016:XST:2939672.2939785}, the loss function for the individual trees of the ensemble is regularized by the quadratic sum of the weights on each leaf multiplied by an ``L2'' regularization hyperparameter, which we fix as $\lambda = 0.01$. The minimum loss reduction for new splittings in a tree is set to $\gamma = 0$. The contribution of each new tree added to the aggregated prediction of the ensemble is scaled by the learning rate, which we set to $\eta = 0.5$.  {\sc XGBoost} hyperparameters were scanned manually around the values used in the study.  One criterion for selection was stability of the results, which do not exhibit strong sensitivity to small variations of these parameters.

We find that certain of the hyperparameters employed by {\sc XGBoost} have extensive (rather than intensive) scaling with the total sample weight.  As such, we find that hyperparameter selections and training outcomes will be less sensitive to variations of the sample size if one normalizes the overall weight to unity in the training.  We do this separately for each of the signal and background classes in order to balance the training datasets.  Physical per-sample weights are used to interpret the results of training.

Any event analyzed by 
the BDT is given a continuous decimal score between 0 and 1, with 1 being most signal-like and 0 being most background-like.  In 
Figure~\ref{fig:score}, we plot the cross section for signal (orange) and background (as labeled) events to be classified 
with a score greater than a given threshold.  Two similar plots made with a different choice of the training 
and evaluation samples are given in Appendix~\ref{app:valid}.  
The far left of the plot (where the signal threshold is zero) corresponds to the case where the BDT results are ignored, and 
yield the cross section for the signal and each background to pass the precuts. The cross section for signal events to 
pass the precuts is $\sim 1.3~\fb$, indicating that, with an integrated luminosity of $\sim 300~\fb^{-1}$, 
$\sim 400$ signal events are expected to pass the precuts and be analyzed by the BDT.  For each of the six backgrounds 
analyzed, the cross section for events to pass the precuts is in all cases $\lesssim 20~\fb$, with the largest cross section 
being for the $\bar \tau \tau jjj$ background.  In particular, note that the $ZZjj$ and $WZjj$ backgrounds have been 
effectively eliminated by the precuts; the analysis of the BDT is largely irrelevant for those backgrounds, as 
$S/B \gg 1$ even before the BDT score is included.

\begin{figure}[htpb]
\centering
\includegraphics[width=0.8\linewidth]{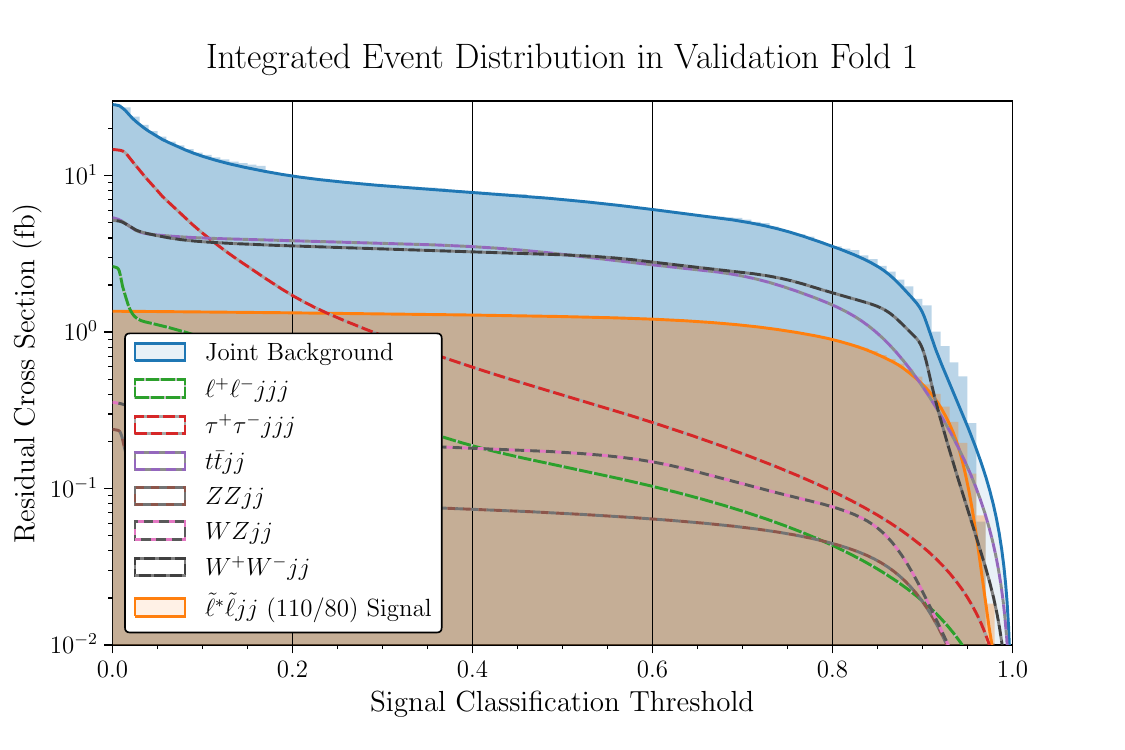} 
\caption{Plot illustrating the residual cross section for signal and background (as labeled)
as a function of the BDT classification score after precuts. Corresponding plots for additional folds of training and evaluation samples can be found in Appendix~\ref{app:valid}.}
\label{fig:score}
\end{figure}

We classify an event as signal-like if its score exceeds a given a threshold, and 
as background-like if otherwise.  Although more complicated methods which do not involve a binary assignment of signal-like 
or background-like status are possible~\cite{Whiteson,Murphy:2019utt,Arganda:2022zbs,Khosa:2022vxb}, 
and likely will provide better sensitivity, 
we will find that this simple method is sufficient to already provide substantial improvement 
in sensitivity.   Moreover, the use of a simple binary assignment will facilitate one of our primary 
goals, which is understanding the physics considerations which underlie the discrimination of signal from 
the various backgrounds.

\subsection{Results}

Although the $\bar \tau \tau jjj$ background is the largest after the precuts, the BDT has little 
difficulty in discriminating this background from signal.  
For this background, as well as $\bar \ell \ell jjj$, the BDT can effectively suppress the background 
with little loss in signal cross section (see Figure~\ref{fig:score}, for a signal classification threshold 
$\gtrsim 0.4$).  
The greatest difficulty which the BDT finds is in 
discriminating the $\bar t t jj$ and $WW jj$ backgrounds from signal.  We may thus expect that the competition between signal 
and the $\bar t t jj$ and $WW jj$ backgrounds will dominate our sensitivity (and associated signal to background ratio).

We see similarly from Figure~\ref{fig:discrimination} 
that the BDT provides a strong ability to discriminate signal from all backgrounds, 
though this discriminating power is weakest for the $\bar t t jj$ and $WWjj$ backgrounds.  
For the $\bar t t j j$ and $WWjj$ cases, although the BDT analysis  
exhibits good discriminating power, the large magnitude of these backgrounds relative to the signal 
cross section implies that, at best, the ratio of signal-to-background is ${\cal O}(1)$.  This 
level of discrimination arises only when the signal classification threshold is high enough (see Figure~\ref{fig:score}) 
that only $\lesssim 150$ signal or background events are selected.

\begin{figure}[htpb]
\centering
\includegraphics[width=0.8\linewidth]{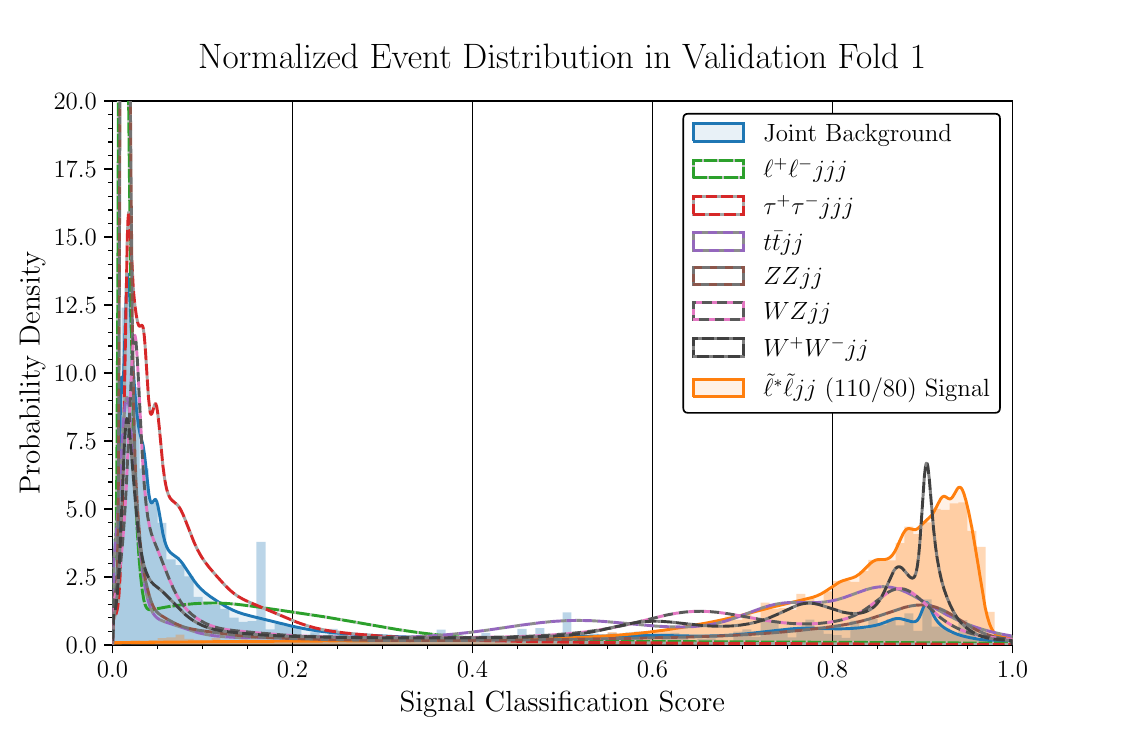} 
\caption{Density plot representing the separation of signal from each background (as labeled) by
the boosted decision tree after precuts. For visual clarity, the densities are separately normalized such that the joint background density is not equivalent to the sum of the individual background densities. Corresponding plots for additional folds of training and evaluation samples can be found in Appendix~\ref{app:valid}.}
\label{fig:discrimination}
\end{figure}

We thus see that optimal performance of the BDT implies a reduction of background down to an event rate comparable to 
signal, with only a factor of $\sim 3$ reduction of the signal event rate.  With $\sim 150$ signal events and $\sim 3$ times 
as many background events, one would expect that a sensitivity of $\gtrsim 6\sigma$ 
could be obtained with an 
$\sim 0.3$ signal to background ratio.  This intuition is confirmed by the results of Figure~\ref{fig:significance}, 
in which we present the number of signal events $S$ (blue), the signal-to-background ratio ($S\div(1+B)$, orange), and 
the signal significance ($\sigma_A$, green)
\footnote{We use the Asimov formula $\sigma_A = \sqrt{2[(S+B)\ln(1+S/B)-S]}$ for the signal significance, which reduces to $\sim S/\sqrt{B}$ for $S/B \ll 1$~\cite{Cowan:2010js}.} as a function of the signal classification threshold, assuming an 
integrated luminosity of $300~\fb^{-1}$.  Note that 
shaded regions show sharp features which arise when the number of simulated events which exceed the signal classification 
threshold is small, and can thus vary dramatically as the threshold increases.  These sharp features  arise when $S$ is 
small and do not represent reliable estimates, as they may be artifacts of the size of the simulated dataset.  
We instead focus on the solid curves, which represent an interpolation 
of the boundary of the shaded regions which smooths out these sharp features.
In Appendix~\ref{app:valid}, the two supplementary plots present analyses with different choices of the sample used for 
training and evaluation.  All three panels are roughly consistent, indicating that our result is robust to variations 
in the training set.  In particular, for a signal classification threshold of $\sim 0.9 $, we find 
that a scenario with $M_{\tilde \ell} = 110~\gev$, $M_X = 80~\gev$ can be detected with 
$\gtrsim 6\sigma$ significance and a signal-to-background ratio of $\sim 0.3$, with $S \sim 100$.
This represents a marked improvement over the results obtained from the cut-based approach used in~\cite{Dutta:2017nqv}, 
in which only $3.0\sigma$ sensitivity (with a signal-to-background ratio of $0.2$) could be obtained with the same 
luminosity  (see Table~\ref{tab:varyprecuts}).\footnote{The signal sensitivity estimated in~\cite{Dutta:2017nqv} is reported as $4.4\sigma$ with a signal-to-background ratio of $0.3$. In this work, we have improved the background modeling and statistics, resulting in a lower signal significance and signal-to-background ratio when implementing the same series of cuts proposed  for intermediate mass gaps.}  Moreover, we see that 
using the cuts imposed in~\cite{Dutta:2017nqv} as precuts, with subsequent signal classification by a BDT, does not 
provide significantly improved sensitivity.  Evidently the cut-based approach used in this earlier analysis results 
in too a large reduction in the number of signal events for optimal signal classification.

\begin{figure}[htpb]
\centering
\includegraphics[width=0.8\linewidth]{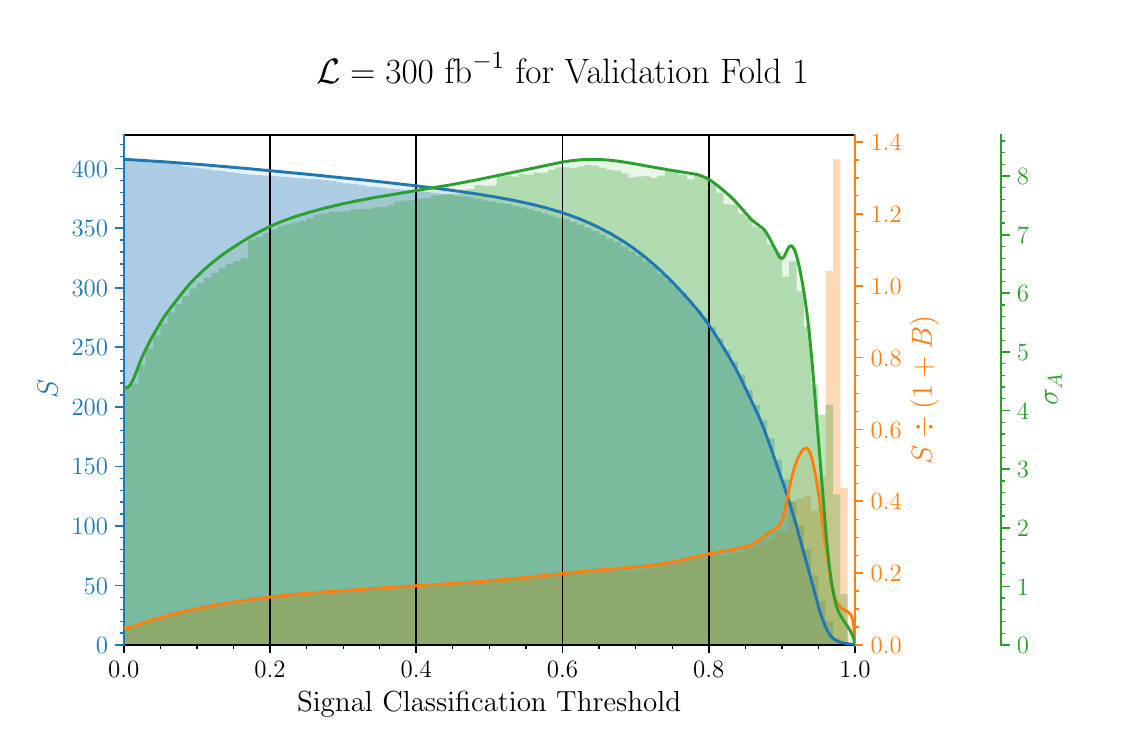}
\caption{The number of signal events ($S$, blue), signal-to-background ration 
($S\div(1+B)$, orange) and signal significance ($\sigma_A$, green) as a function of the 
signal classification threshold. Each curve represents an interpolation which smooths out the sharp 
features in the shaded region, which arise from small event numbers. Corresponding plots for additional folds of training and evaluation samples can be found in Appendix~\ref{app:valid}.}
\label{fig:significance}
\end{figure}

It is important to include the effects of systematic uncertainties on the sensitivity of 
this search.  We do not provide a quantitative estimate, as the systematic uncertainties need not be 
Gaussian.  We instead use a qualitative assessment: provided the signal-to-background ratio is substantially 
larger than the estimated systematic uncertainties, one expects that an underestimated background cannot 
duplicate the effects of a signal discovered with high statistical significance.
An estimate by ATLAS of  the background of a 
slepton search suggests systematic uncertainties at the level of $\sim 17\%$~\cite{ATLAS:2022hbt,ATLAS:2019lng}.  
As the signal-to-background ratio obtained in this analysis is substantially larger than this ($S/B \sim 0.3$), one expects to be able 
to detect the presence of signal with sufficient luminosity.

Since the main work of the BDT (beyond the precuts, whose intuition we understand) is in rejecting the 
$\bar t t jj$ and $WWjj$ backgrounds, we can ask which kinematic variables the BDT used in that analysis.  These results 
are shown in Figure~\ref{fig:VariableImportance} which indicate which kinematic variables dominate the ``total gain''~\footnote{This measure of relative importance compares the summed reduction in the loss function that is
attributable to leaf splittings associated with each available feature.  It is sensitive to
how often a feature is used across nodes, how many events pass through each such
node, and the weights carried by those events.} 
when a BDT is trained only against a single background.  This figure thus gives an indication 
of which variables are most useful in discriminating signal from a particular background.  
In particular, the most important kinematic variable for rejecting the $ttjj$ and $WWjj$ backgrounds 
is $M_{\rm T2}^{100}$, which is defined, as in Ref.~\cite{Lester:1999tx}, as the minimal mass of a parent particle  under the 
assumption that the parent particles are pair produced and each decays to a lepton and an invisible particle 
with a mass of $100\gev$.  Note that this variable was also used in the slepton search analysis presented in 
Ref.~\cite{ATLAS:2022hbt}, though with a different purpose.

\begin{figure}[htpb]
\centering
\includegraphics[width=0.475\linewidth]{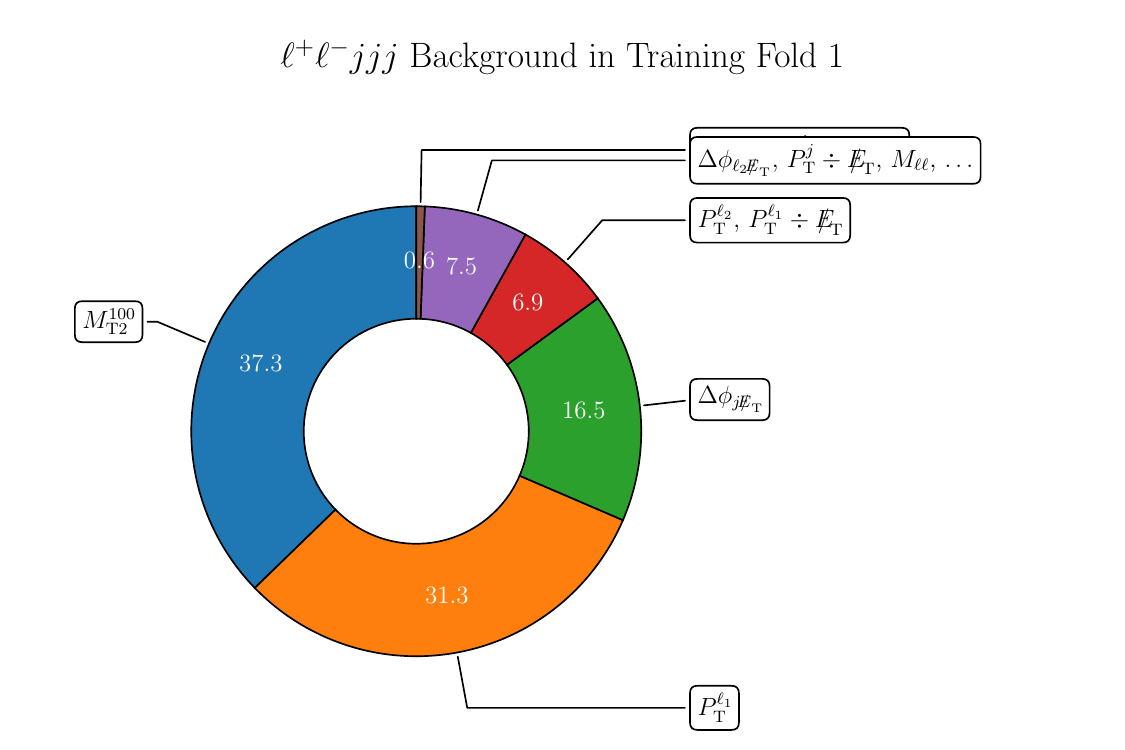} \hspace{6pt}
\includegraphics[width=0.475\linewidth]{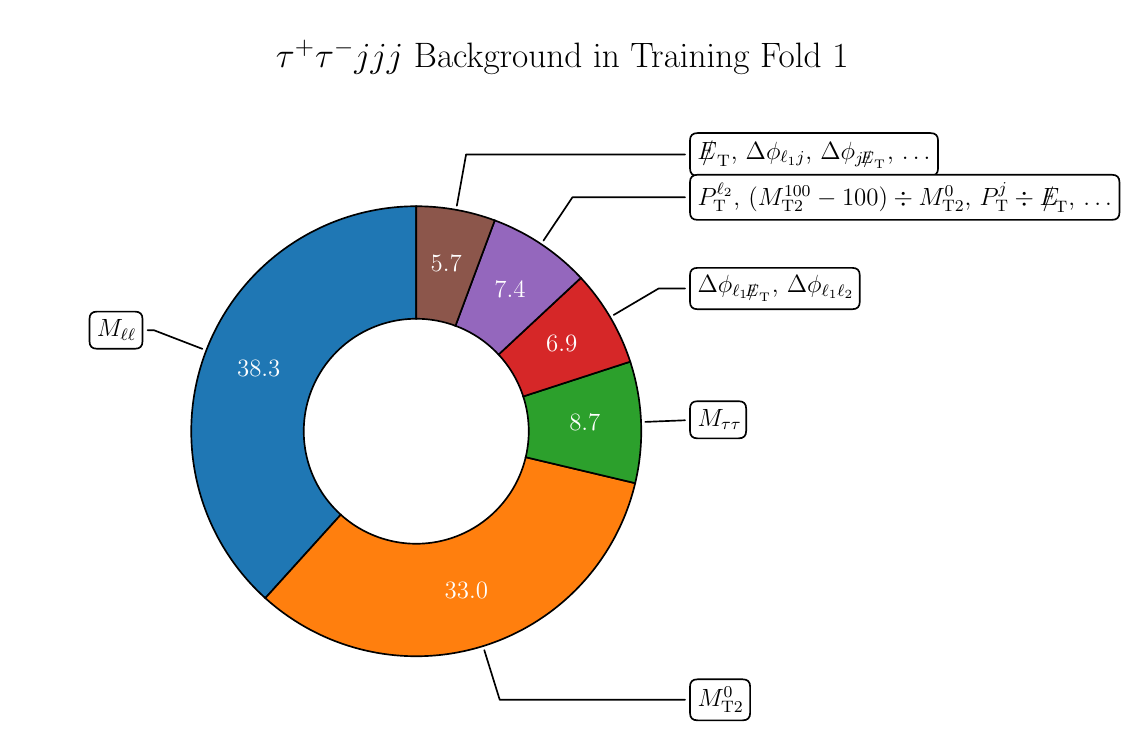} \\
\includegraphics[width=0.475\linewidth]{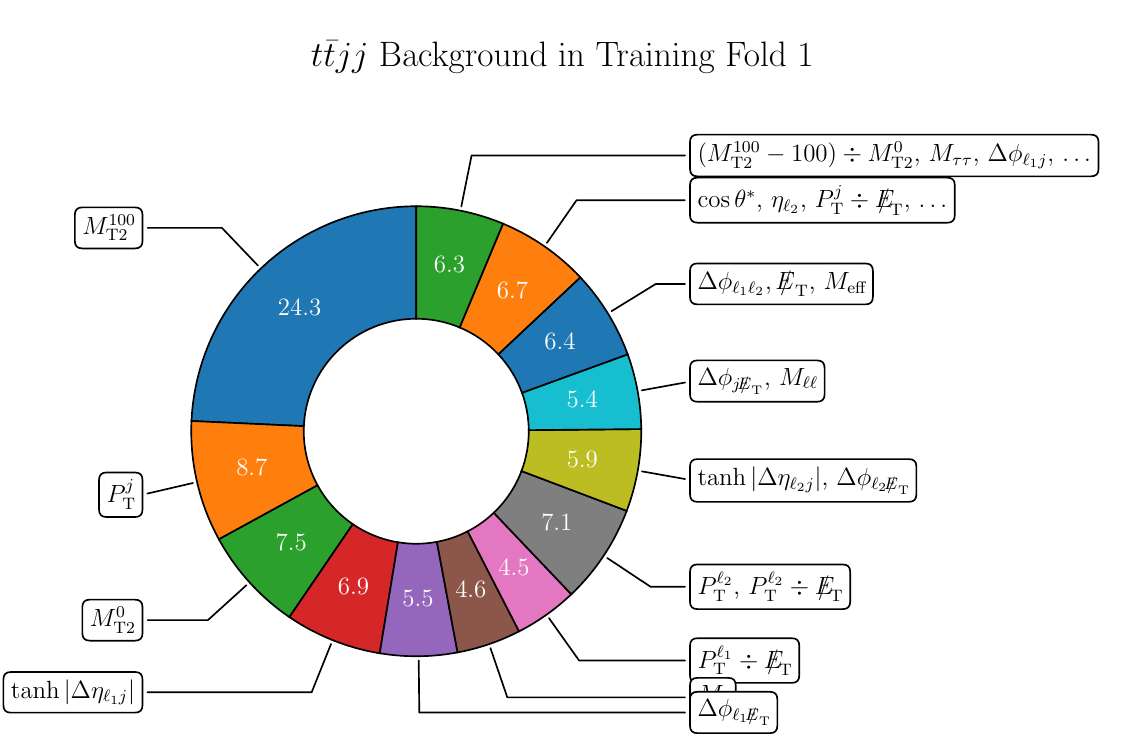} \hspace{6pt}
\includegraphics[width=0.475\linewidth]{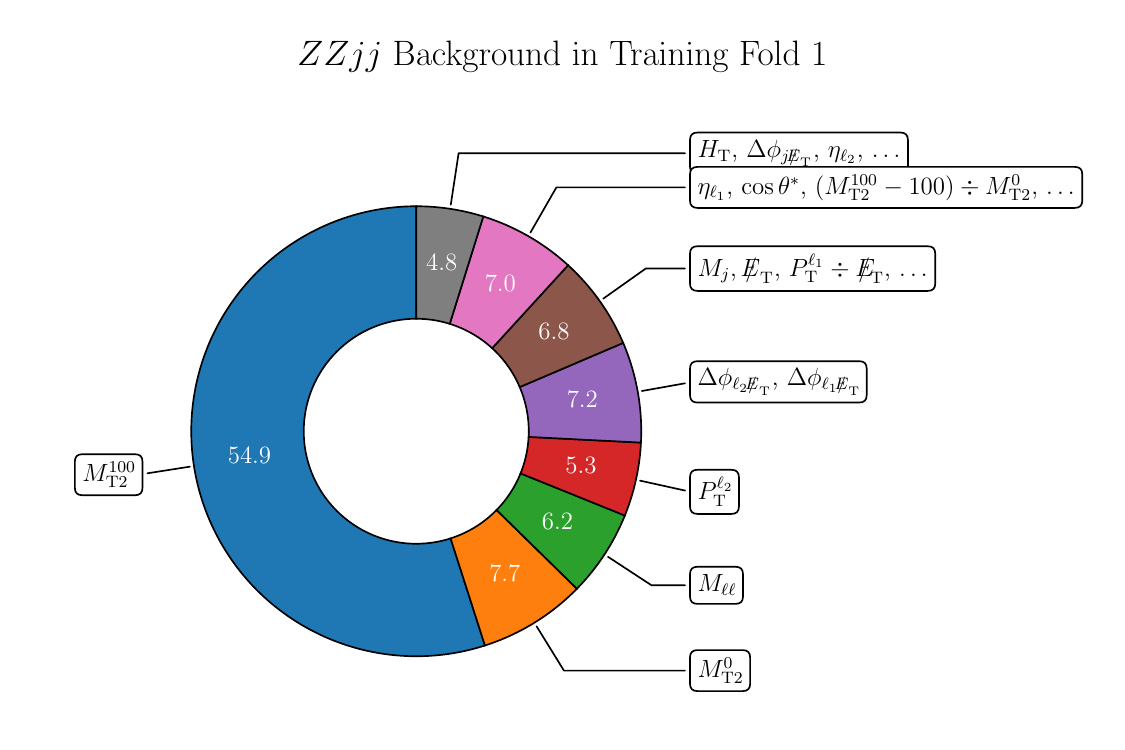} \\
\includegraphics[width=0.475\linewidth]{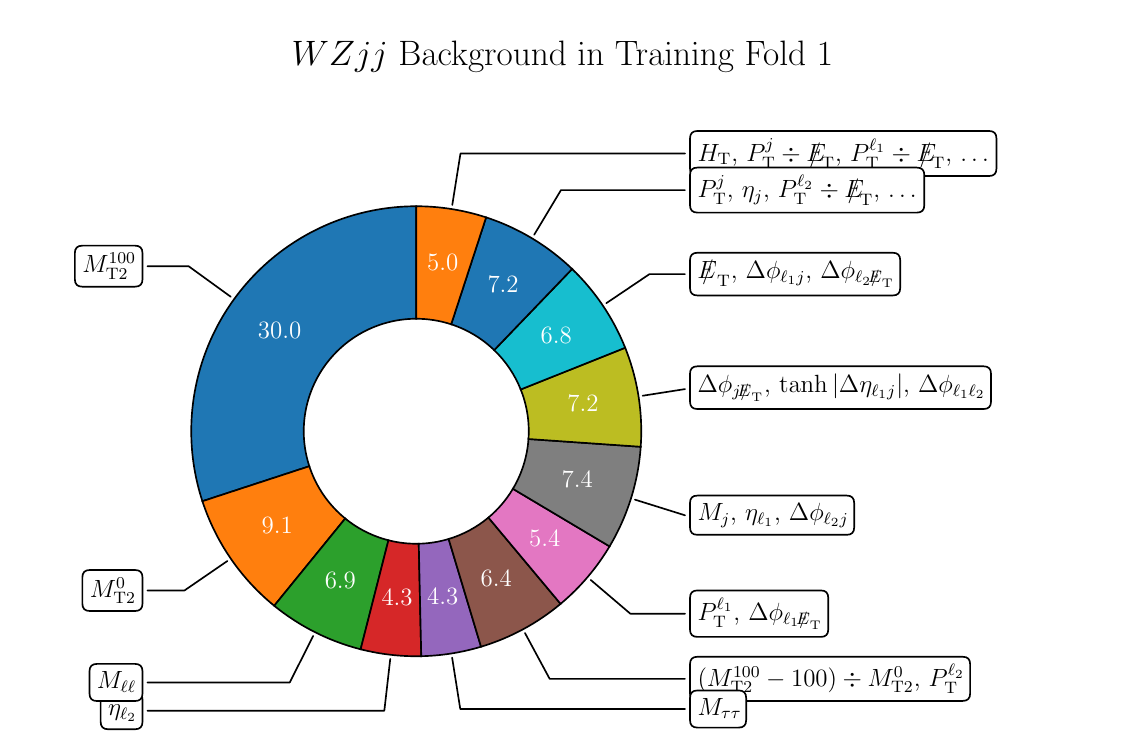} \hspace{6pt}
\includegraphics[width=0.475\linewidth]{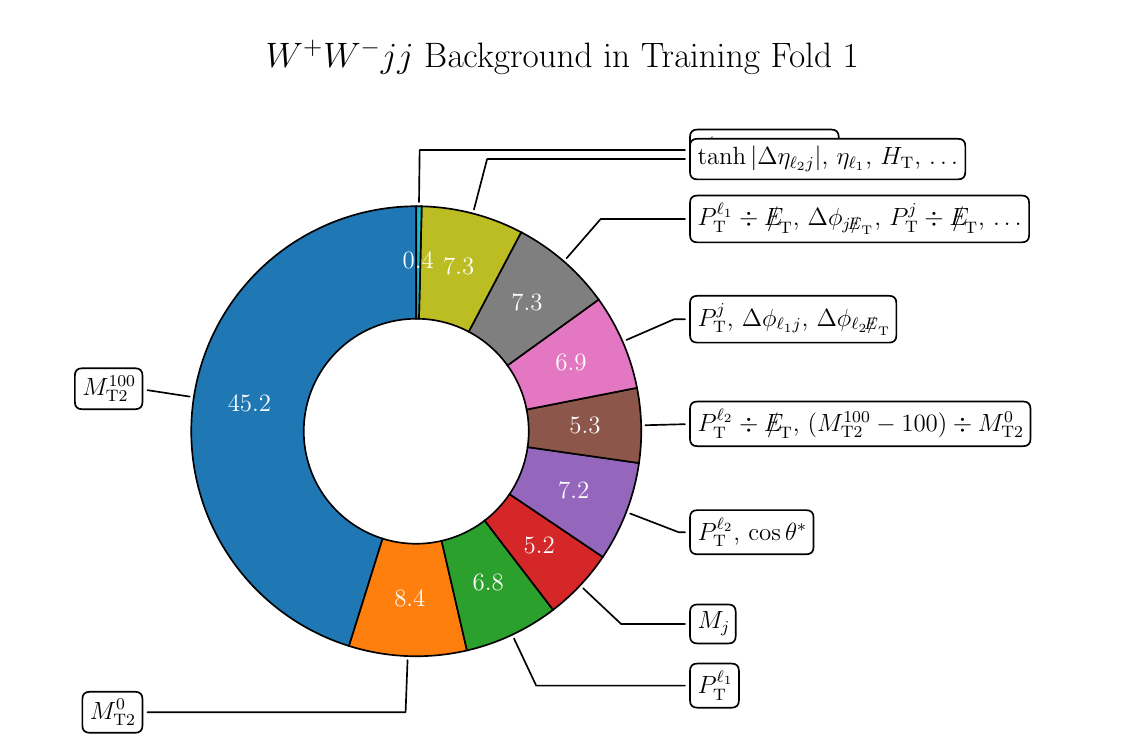}
\caption{Plots indicating the relative importance of various kinematic
features to the separation of signal from background by
the boosted decision tree after level 1 pre-selections, respectively
(left-to-right then top-to-bottom) for $\ell \ell jjj$, $\tau \tau jjj$,
$t\bar{t} jj$, $ZZjj$, $WZjj$, and $WWjj$.
Note that, in each panel, results are shown for a BDT which is trained only 
with signal and the particular background process in question.
}
\label{fig:VariableImportance}
\end{figure}

The distributions of $M_{\rm T2}^{100}$ 
for signal and $\bar t t jj$ and $WWjj$ background events (after precuts) are given in Figure~\ref{fig:Distributions}.
We can see from 
these distributions that $M_{\rm T2}^{100}$ contributes significant ability to discriminate between signal and background.  
In particular, background events are biased towards higher values of $M_{\rm T2}^{100}$, since for background events, the 
invisible particles are neutrinos which are much lighter than the ansatz of $100 \gev$.

\begin{figure}[htpb]
\centering
\includegraphics[width=0.8\linewidth]{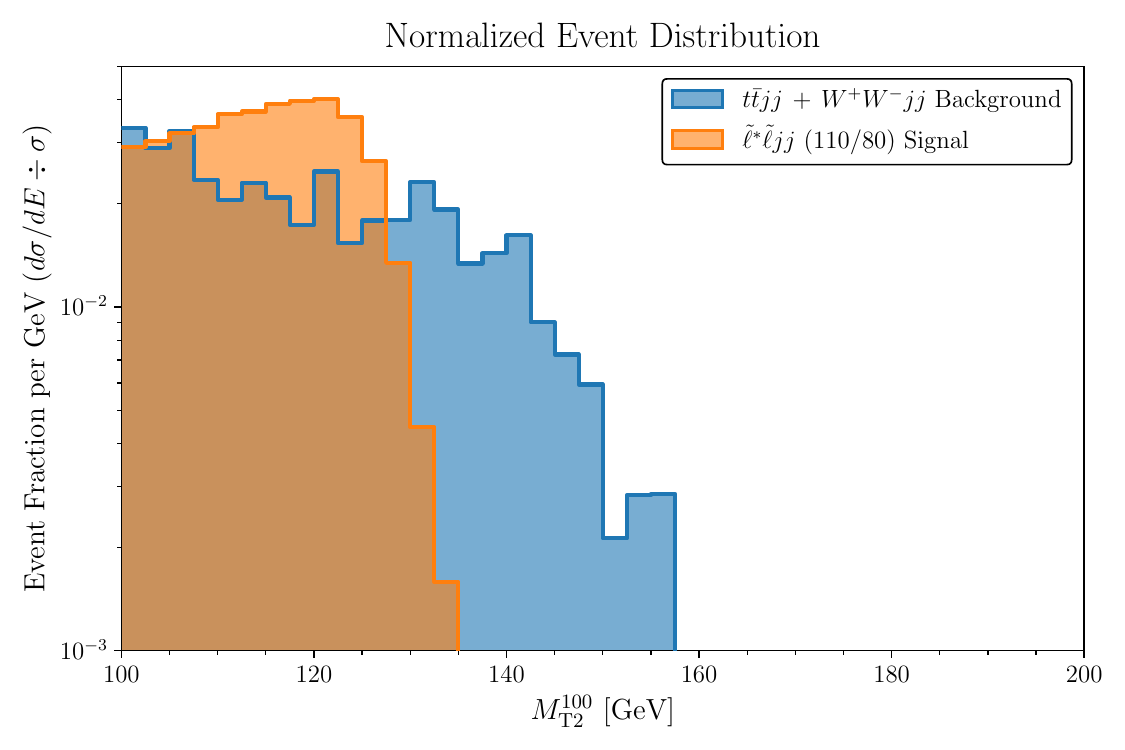}
\caption{The distribution of $M_{T2}^{100}$ for events passing the precuts for 
signal (orange), as well as the combined $ttjj$ and $WWjj$ backgrounds (blue).  
}
\label{fig:Distributions}
\end{figure}

However, these distributions do not provide a clear sense of how $M_{\rm T2}^{100}$ is correlated with 
other variables, which is the type of information one would expect to be learned by a BDT.
To consider this question, we assume that the only relevant background processes are 
$ttjj$ and $WWjj$.  That is, we ignore the $WZjj$ and $ZZjj$ background classes (which are essentially 
eliminated by the precuts) and the $\mu \mu jjj$ and $\tau \tau jjj$ background classes (which are 
easily eliminated by the BDT).  After the precuts, we see from Figure~\ref{fig:discrimination} that 
we are left with $\sim 400$ signal events, and $\sim 3300$ background, roughly evenly split between 
$ttjj$ and $WWjj$.  We then consider a simple cut in which we accept only events with 
$M_{\rm T2}^{100} < 130 \gev$.  This cut admits almost all signal events, but reduces each class of background events by 
roughly 1/3.  This simple cut-and-count analysis would yield  a signal-to-background ratio of $S/B \sim 0.18$, 
with a signal significance of $\sim 8.5\sigma$. 
Indeed, we see that a 1/3 reduction in background, with a small reduction in signal, is roughly 
what the BDT achieves with a signal classification threshold set at $\sim 0.6$ (see Figure~\ref{fig:score}), yielding 
approximately the significance and signal-to-background ratio we estimated with a cut-and-count analysis (see Figure~\ref{fig:significance}).
The deeper correlations found by the BDT (at least in regard 
to the $ttjj$ and $WWjj$ backgrounds) lead to an improvement in the signal-to-background ratio of roughly a 
factor of 2 as the signal classification threshold is raised to $>0.9$.  
Note that this improvement in the signal-to-background ratio is essential for overcoming 
systematic uncertainties, as described above.

Since the imposition of a simple cut such as $M_{\rm T2}^{100} < 130 \gev$ seems to replicate at least part of 
the BDT analysis, one might ask if imposing this as a precut could improve the performance of the 
BDT, by allowing it to focus on less obvious correlations.  To test this possibility, we ran a second analysis in 
which  $M_{\rm T2}^{100} < 130 \gev$ was imposed as a additional precut.  The results are fairly similar to analysis without 
this additional precut (see Table~\ref{tab:varyprecuts}).  
It appears that, because the $ttjj$ and $WWjj$ backgrounds are not excessively large 
compared to the other backgrounds and are within an order of magnitude of the signal, 
after applying the precuts described in Section~\ref{sct:bdttraining}, 
the BDT was not overly focused on reducing them, and thus did not gain significantly in performance when the  $ttjj$ and $WWjj$ backgrounds 
were reduced through an additional precut.

\begin{table}[ht]
\centering
\bgroup
\def\arraystretch{1.5}
\begin{tabular}{|c|c|c|c|c|c|c|}
\hline
\multirow{2}{*}{Event Selection} &
\multicolumn{2}{|c|}{\quad This work \quad} &
\multicolumn{2}{|c|}{\quad This work $+$ $M_{\rm T2}^{100} < 130 \, \gev$ \quad} &
\multicolumn{2}{|c|}{\quad Intermediate mass gaps in~\cite{Dutta:2017nqv} \quad}
\\ 
\cline{2-7}
& \quad Precuts \quad & \quad  BDT  \quad  & \quad Precuts \quad & \quad  BDT \quad  & \quad Precuts \quad & \quad  BDT \quad  \\
\hline
\quad BDT signal threshold \quad & - & \quad $0.47-0.90$ \quad & - & $0.34-0.88$ & - & $0.00-0.71$ \\
\quad Events at $\mathcal{L} = 300 \, \ifb$ \quad & $413$ & \quad $107-386$ \quad & $405$ & $121-387$ & $48$ & $26-51$ \\
$S \div (1+B)$ & $0.05$ & \quad $0.15-0.36$ \quad & $0.06$ & $0.15-0.31$ & $0.20$ & $0.18-0.35$ \\
$\sigma_A$ & $4.4$ & \quad $5.8-8.2$ \quad & $4.8$ & $5.6-8.2$ & $3.0$ & $2.5-3.7$ \\
\hline
\end{tabular}
\egroup
\caption{Summary of results for our benchmark mass spectrum (i.e. $M_{\tilde \ell} = 110 \, \gev$ and $M_{\tilde \ell} - M_X = 30 \, \gev$) after implementing different event selections. For each set of precuts, we show the number of signal events, the signal to background ratio and the signal sensitivity both before and after application of the BDT. When only considering the precuts, we average the respective metrics between the 3 folds of the data. In contrast, the respective BDT signal score thresholds are optimized to show as broad a range of $\sigma_A$ possible between the different folds while maintaining $S \div (1+B) \geq 0.15$. The first event selection summarizes the the results of the event selection with precuts described in Section~\ref{sct:bdttraining}. The second event selection adds a precut on $M_{\rm T2}^{100}$ and the third event selection implements precuts corresponding to that which is implemented for the intermediate mass gap scenarios in previous work~\cite{Dutta:2017nqv}. Note the average number of events after precuts between all 3 folds of the data in this final event selection is smaller that the upper end of the range of numbers of signal events after application of the BDT because of the variation between folds at a signal score threshold of 0.00, which is equivalent to only applying the precuts.
}
\label{tab:varyprecuts}
\end{table}

\begin{table}[ht]
\centering
\bgroup
\def\arraystretch{1.5}
\begin{tabular}{|c|c|c|c|c|c|c|}
\hline
\multirow{2}{*}{} &
\multicolumn{2}{|c|}{$M_{\tilde \ell} = 110 \, \gev$} &
\multicolumn{2}{|c|}{$M_{\tilde \ell} = 160 \, \gev$} &
\multicolumn{2}{|c|}{$M_{\tilde \ell} = 300 \, \gev$}
\\ 
\cline{2-7}
& \quad Benchmark \quad & \quad Retrained  \quad  & \quad Benchmark  \quad & \quad Retrained   \quad  & \quad Benchmark  \quad & \quad Retrained   \quad  \\
\hline
\multirow{4}{*}{$\Delta M = 30 \, \gev$} & $0.47-0.90$ & - & $0.91-0.92$ & $0.90-0.94$ & - & - \\
 & $107-386$ & - & $33-55$ & $13-43$ & - & - \\
 & $0.15-0.36$ & - & $0.18-0.32$ & $0.15-0.39$ & $<0.15$ & $<0.15$ \\
 & $5.8-8.2$ & - & $2.5-3.1$ & $1.8-2.9$ & - & - \\
\hline
\multirow{4}{*}{$\Delta M = 40 \, \gev$} & $0.47-0.91$ & $0.23-0.87$ & - & $0.90-0.92$ & - & - \\
& $68-375$ & $163-496$ & - & $24-49$ & - & - \\
 & $0.15-0.36$ & $0.15-0.35$ & $<0.15$ & $0.15-0.25$ & $<0.15$ & $<0.15$ \\
 & $4.2-7.9$ & $6.7-9.1$ & - & $1.9-3.0$ & - & -\\
\hline
\multirow{4}{*}{$\Delta M = 50 \, \gev$} & $0.59-0.90$ & $0.13-0.88$ & - & $0.79-0.93$ & - & - \\
& $67-320$ & $172-575$ & - & $38-156$ & - & -\\
 & $0.15-0.23$ & $0.15-0.42$ & $<0.15$ & $0.15-0.35$ & $<0.15$ & $<0.15$  \\
 & $3.7-7.3$ & $7.6-10.7$ & - & $3.0-5.4$ & - & - \\
\hline
\multirow{4}{*}{$\Delta M = 60 \, \gev$} & $0.62-0.90$ & $0.05-0.90$ & - & $0.54-0.96$ & - & $0.96$ \\
& $60-295$ & $252-691$ & -& $21-273$ & - & $14-16$ \\
 & $0.15-0.21$ & $0.15-1.01$ & $<0.15$ & $0.15-1.16$ & $<0.15$ & $0.15-0.26$ \\
 & $3.3-6.9$ & $9.9-14.9$ & - & $3.6-7.9$ & - & $1.5-1.9$ \\
\hline
\end{tabular}
\egroup
\caption{Summary of results for a variety of different mass spectra. For each combination of scalar mass $M_{\tilde \ell}$ and mass splitting $\Delta M = M_{\tilde \ell} - M_X$, we show (from top to bottom) the optimal signal score threshold for the BDT, the number of signal events expected at $\mathcal{L} = 300 \, \ifb$,  $S \div (1+B)$ and $\sigma_A$. For each case, we test the BDT trained on the benchmark spectrum (i.e. $M_{\tilde \ell} = 110 \, \gev$ and $\Delta M = 30 \, \gev$) and the BDT retrained specifically for each spectrum. The respective BDT  signal score thresholds are optimized to show as broad a range of  $\sigma_A$ possible while maintaining $S \div (1+B) \geq 0.15$.}
\label{tab:varybench}
\end{table}

\subsection{Other Benchmarks}
\label{sct:OtherBenchmarks}

Until now, we have considered a single benchmark parameter point: 
$M_{\tilde \ell} = 110~\gev$, $M_X = 80~\gev$.  Here we consider the 
sensitivity of an LHC search with $300~\fb^{-1}$ integrated luminosity 
to models with different choices for the masses.  In Table~\ref{tab:varybench}, 
we consider $M_{\tilde \ell} = 110, 160, 300~\gev$, with
$M_{\tilde \ell} - M_X = 30, 40, 50, 60~\gev$.  In each case, we either use 
BDT trained against the original benchmark model, or against the model to be 
analyzed (in all cases, the original precuts are used).  For each analysis, 
we present an optimal range for the signal classification thresholds, and the range of the 
number of signal events, signal to background ratio, and signal significance as 
one varies the signal classification threshold within its optimal range.  We only present 
results for $S \div (1+B) \geq 0.15$, in order to ensure that the analysis is robust against likely systematic 
uncertainties~\cite{ATLAS:2022hbt,ATLAS:2019lng}.

For cases with $M_{\tilde \ell} = 110~\gev$ and larger mass gaps, $M_{\tilde \ell} - M_X > 30 ~\gev$, we see that the BDT trained on the benchmark spectrum, $M_{\tilde \ell} = 110~\gev$ and $M_{\tilde \ell} - M_X = 30 ~\gev$, maintains robust sensitivity to scalar muon production even with some degradation as the mass gaps become larger. Alternatively, when training the BDT individually for each benchmark, we see that the sensitivity is enhanced at larger mass splittings due to the improved separation of signal from background in the kinematic distributions. Although mass gaps $M_{\tilde \ell} - M_X = 50, 60~\gev$ are currently excluded  for $M_{\tilde \ell} = 110~\gev$, the signal significance we project for the BDT analysis of these cases is considerably larger than in the most stringent LHC constraints from Ref.~\cite{ATLAS:2022hbt}, which exclude mass splittings of $\sim 50~\gev$ at $\sim 2\sigma$.  Similar dependencies of the sensitivity in the BDT analysis on the mass splitting can be seen for $M_{\tilde \ell} = 160~\gev$, however the BDT trained on the benchmark mass spectrum is only sensitive to the scenario with the same mass splitting of $M_{\tilde \ell} - M_X = 30 ~\gev$. When the BDT is retrained for each mass spectrum, we see the BDT could be sensitive to the full range of mass splittings from $30 ~\gev$ to $60 ~\gev$, none of which are currently constrained by LHC searches.

For $M_{\tilde \ell} = 300~\gev$, we never find $S \div (1+B)$ much larger than $0.15$, implying that 
a BDT search using these precuts might be difficult even at higher luminosity.  But it is worth 
noting that, for such a large lepton partner mass, the cross section for signal events to pass the 
precuts is much smaller than each of the backgrounds, with the total background well over two orders 
of magnitude larger than the signal, after imposing the precuts.  The precuts were imposed precisely 
to eliminate this hierarchy, and we see that the precuts chosen to be adequate for 
$M_{\tilde \ell} = 110~\gev$ do not achieve this purpose if the lepton partner mass is much heavier.  

Better prospects would likely lie in developing more aggressive precuts to study this higher mass range 
(likely using higher luminosity).  
But training a BDT after the imposition of more aggressive precuts would require a much larger sample of simulated events. 
The difficulty lies in ensuring that the training data set adequately samples the tails of the kinematic distributions, 
where it may be most difficult to distinguish signal events from background events.  As one imposes more aggressive 
precuts, one will find that fewer events in the training sample will pass the precuts.  As a result, one may find that 
at the tails of the kinematic distributions, training is dominated by only a few events, leading to unreliable BDT performance 
which is biased by the indiosyncracies of the training dataset.  To apply precuts designed for much higher luminosities while 
avoiding this difficulty, one should generate a much more extensive set of simulated data.  Although that is beyond the scope of 
this work, it would be an interesting topic of future study.

\subsection{Restricted Training} \label{sct:restrictedtraining}

In this subsection we assess the question of how much classification power remains if one restricts the set of available variables for training.  Specifically, we start from the precuts described in Section~\ref{sct:bdttraining}, while passing only a portion of the variables in Table~\ref{tab:variables}
to the BDT.  Specifically, we test restricted training for just the most effective variables (as indicated by averaging feature importance to total gain over the three folds for training of signal against the joint background) and also for a selection of ``low-level'' variables (including ${/\!\!\!\!E}_{\rm T}$, $H_{\rm T}$, the transverse momentum values, as well as the azimuthal and rapidity separation variables). 

\begin{figure}[htpb]
\centering
\includegraphics[width=0.8\linewidth]{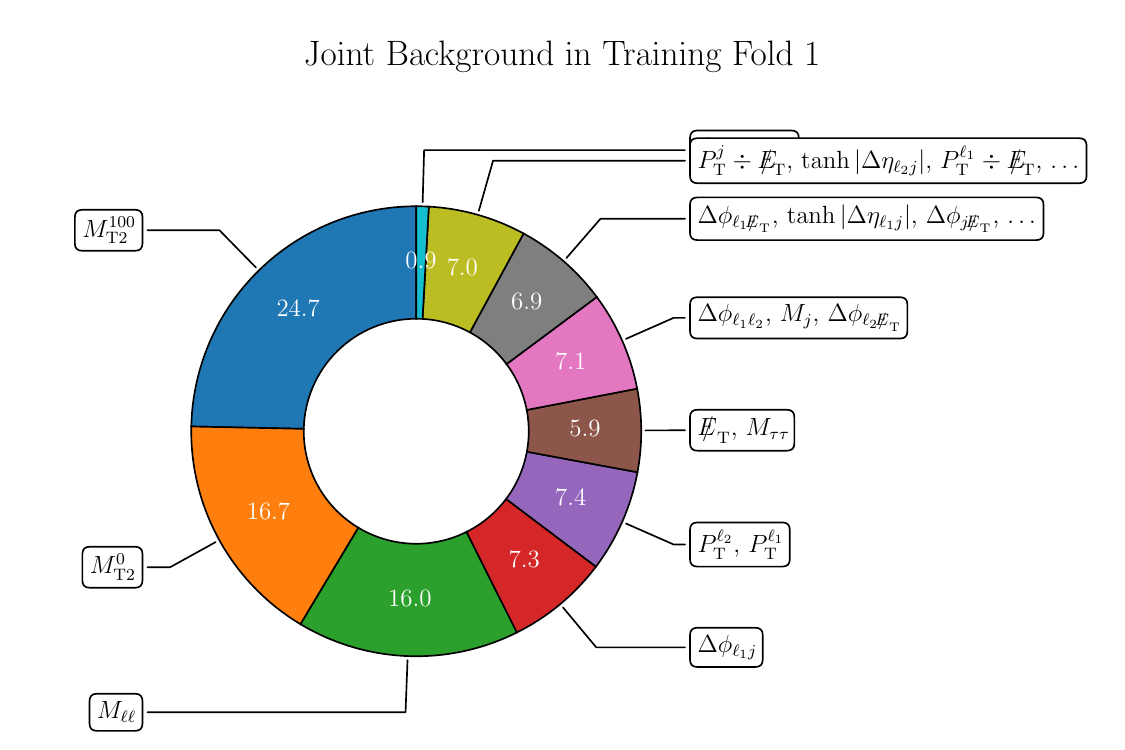}
\caption{Plot indicating the relative importance of various kinematic
features to the separation of signal from the joint (all source) background.
This training occurs application of the Section~\ref{sct:bdttraining} precuts.}
\label{fig:jointimportance}
\end{figure}

Figure~\ref{fig:jointimportance} illustrates the feature importance (for fold~1)
with joint training against all backgrounds after the application of precuts.
We use these results (including folds~2 and~3) to select the top performing variables for restricted training.
The top performing variables for joint training after precuts are all high-level variables associated with reconstructed masses (or lower mass bounds), namely $M_{\rm T2}^{100}$, $M_{\rm T2}^0$, and $M_{\ell\ell}$, in that order.  These three features account for more than half of the total gain, and are important across all three training folds. By contrast, no other feature has an average importance above about 6\%, and
more variability is observed across the folds.  For these reasons, we include just these three features in the restricted training for top-performing variables.  This has the added benefit of precluding any direct overlap with the restricted training on low-level features.

We find that the best results are indeed obtained using the full
ensemble of both high- and low-level observables.  
The peak signal-to-background ratio (using the interpolated measure) is reduced
by about 40\% for each of the restricted trainings relative to
the case of training on all variables.
In terms of statistical significance, the high-level
variables almost match the performance of the full set,
whereas the low-level observables give up
about 40\% in this regard as well.
Since the success of this analysis hinges critically
on elevating the signal-to-background ratio above the
systematic uncertainty noise floor, it becomes
all the more important to supplement
the top-performing high-level observables
with the larger number of weaker correlations provided
by the low-level observables.

\section{Conventional Wisdom for BDTs\label{sct:conventionalwisdom}}

There are several lessons that have been learned during the course of
this study that would seem to represent a type of conventional wisdom which is
substantially transferable in other similar applications.

The approach described in Section~\ref{sct:motivation} works reasonably well ``out of the box,'' but we have observed that
it is possible to achieve significantly better separation between signal and background using a hybrid technique that
involves significant human input in conjunction with training of the BDT.
By contrast, the emerging understanding in the deep-learning community
is that one should generally just ``get out of the way'' and let the training proceed with minimal supervision and data curation. However, that intuition is apparently
less applicable in the context of more basic types of machine learning applications such as 
boosted decision trees.  The upside is that this additional investment of effort
comes with a very substantially expanded and auditable inventory of precisely
what and how the machine was learning, while still typically accessing a level of feature
separation that meaningfully exceeds what is available with manual feature selections after maximal effort.

In a similar vein, the BDT seems to take maximal advantage of high-level features
that have been constructed to efficiently encode physics features that are expected to be relevant.
Again, by contrast, deep-learning approaches are known to work exceptionally well on raw low-level
data, and extensive human curation is generally not necessary, or may even be detrimental. 
This is because a neural network of sufficient complexity can essentially mimic any mathematical
transformation, and training will drive the development of this function in an optimized way.
By contrast, shallow decision trees are always splitting on single features, and they tend to
operate most effectively when the available features are preconfigured to very densely encode
the most relevant parameterization of available information.  For example, we generally find
that reconstructed masses and composite constructions like $M_{\rm T2}$ or $\cos\theta^*$ pack
much more punch than isolated kinematic features.  Even simple operations, like taking
differences or ratios of scales or angular coordinates that are expected to have more
discriminating power as relative values than absolute values can increase efficiency.  

Additionally, a very deep layer of background will be quite difficult for the BDT to ``see through,''
and it is helpful to remove ``obvious'' backgrounds prior to engaging the BDT.  Along these
lines, it is often the case that the most dominant backgrounds can be the simplest to defeat,
and it seems that this should always be done up front when it is possible.  For example, the
cross section of single $Z$ events is extremely large, and its magnitude will typically overwhelm
any search for final states that feature opposite-sign same-flavor dileptons.  However, the vast
majority of these dilepton events will cluster around the resonant $Z$ mass, and the residual
cross section can be reduced dramatically with a simple window cut exclusion.  Of course, the
BDT can learn such a cut on its own.  However, it seems to become ``exhausted'' in the process,
and it becomes much harder for it to see the more subtle separations which are subsequently
critical to achieving a successfully refined training if the dominant fraction of the input
event weights are telling the BDT to pay attention to the mass window as a first priority.
Not only should ``obvious''
discriminants be applied as manual precuts, it is advisable that the hardness of these
cuts be selected so that the residual cross section associated with each type of background
are brought approximately into parity. 

On a related note, for practical reasons described previously,
the starting cross section of signal events is always
much smaller than that of relevant backgrounds.
We find that the behavior of the BDT will be much more uniform and predictable
if one separately normalizes the sum of signal and background weights to unity
prior to training.  This is because several of the hyperparameters associated with
the BDT implementation, in particular those associated with regulation of the objective,
have extensive (rather than intensive) scaling properties.  This normalization
helps to ensure that intuition developed regarding useful settings of these parameters
has improved cross-applicability to other contexts.  In fact, useful hyperparameter
values generically tend to be ``order one'' in this normalization.  Of course,
one must then rescale back to physical cross sections when making predictions
for rates in a real world experimental environment.

Since we are only really interested here in classification of categories that are somewhat difficult to
separate by elementary means, the BDT will presumably need to access rather narrow
and subtle features in order to train successfully.
Likewise, since we are only really interested in enhancing the visibility of extremely rare
processes, it is necessary to filter away the vast majority of competing background processes.
These realities would appear to usher in a fundamental dilemma common to all relevant
classification problems in collider physics.  This is that the training is only successful
in the regime where it has become statistically limited, and therefore less reliable.
In other words, ``harder cuts'' leave fewer events, and it becomes successively
more difficult to validate that the selected cuts are generalizable.
This concern must be balanced against the priorities identified previously, such
that any precuts will not preclude effective training.  Specifically, there must
be a statistically significant sampling of events on hand for the BDT to process.
Of course, if it is possible to reduce all relevant background via manual selections,
then it may not be necessary to further employ machine learning at all.  However,
overly aggressive precuts can additionally prevent the BDT from acting in a more
surgical manner, and potentially achieving comparable background reductions while
retaining a greater density of signal. 
Although the BDT is technically classifying rather than cutting, the same dilemma
ultimately applies to its training process as well,
and one often in practice contemplates an effective selection cut
on some threshold of the resulting classification score.

\section{Conclusion} \label{sct:conclusion}

We have considered the LHC sensitivity to a scenario of scalar lepton partner pair-production, with each 
partner decaying to a muon and an invisible particle with an intermediate mass splitting
($M_{\tilde \ell} - M_X \sim 30~\gev$).  This is a scenario for which current LHC sensitivity has 
not exceeded LEP, 
owing to the large electroweak background.
We have used an analysis based on a boosted decision tree (BDT), and have found that with $300~\fb^{-1}$ luminosity, 
the LHC could achieve $\gtrsim 5\sigma$ sensitivity to currently allowed models ($M_{\tilde \ell} \sim 110~\gev$), 
with a signal to 
background ratio $\gtrsim 0.3$.  
The BDT analysis could also exclude spectra with larger mass gaps $M_{\tilde \ell} - M_X \sim 50~\gev$ and the same scalar mass with a significance $\gsim 4$ times larger than the most recent LHC analysis~\cite{ATLAS:2022hbt}. 
With the same luminosity, LHC could exclude models with $M_{\tilde \ell}$ as large as
$160~\gev$.  But for a larger mass splitting ($\sim 60~\gev$), the LHC could provide $> 5\sigma$ evidence for 
models with $M_{\tilde \ell} \sim 160~\gev$ (also allowed by analyses of current data).
The projected sensitivity of cut-based analyses in previous theoretical studies suggested that LHC could only be sensitive to models with mass splittings $\sim 30-60~\gev$ for scalar masses not much large than $\sim 110~\gev$~\cite{Dutta:2017nqv}.
Thus, a BDT analysis
in scalar lepton partner searches could definitively probe realizations of the MSSM in which scalar muons with $M_{\tilde \ell} \lsim 150~\gev$ \cite{Fukushima:2014yia} mediate interactions that can both deplete the relic density through dark matter annihilation and account for the anomalous magnetic moment of the muon.

The most difficult backgrounds to discriminate are $\bar t t~{\rm and}~WW~+$~jets.  The kinematic variable relied on most heavily 
by the BDT in discriminating signal from these background is $M_{\rm T2}^{100}$.  Nevertheless, the BDT makes good use 
of less obvious variables to simultaneously suppress the $\mu \mu~+$~jets and $\tau \tau~+$~jets backgrounds while further reducing $\bar t t~{\rm and}~WW~+$~jets, producing an optimal signal-to-background ratio which is a factor of $\sim 6$ better than one would get by simply imposing an additional cut on $M_{\rm T2}^{100}$. 

In performing this analysis, we have learned several lessons regarding the application of BDTs to LHC analyses which 
may be applied to other searches.  In particular, we found 
that the BDT is much more effective if obvious precuts are applied 
which reduce each surviving background to roughly equal event numbers, so that the BDT is not overly dominated by any 
one background class. Essentially, there is little gained by having a BDT relearn things we already know.  On the 
other hand, while the imposition of precuts which reduce very large backgrounds improves the BDT performance, the 
application of even obvious precuts which have the effect of removing backgrounds which are only comparable to the 
signal has little effect on the BDT's performance.

There any many avenues open for further study.  
In particular, it would be interesting to explore the optimal tradeoff between using precuts to reduce backgrounds 
before the application of a BDT, as opposed to simply allowing a BDT to analyze as large a data set as possible.  
It would also be interesting to investigate other approaches to the problem of backgrounds with widely disparate cross sections, 
such as changes to the BDT hyperparameters. 

We have found that it is difficult to properly train a BDT if the most difficult backgrounds to remove are swamped by 
larger (though more tractable) backgrounds.  The essential problem is that the simulation data which one uses for 
training are necessarily only a fraction of the vast LHC dataset.  A subleading background may be undersampled, 
leading a BDT to focus its training on quirks of the training set, rather than robust features. But this is a more 
general danger which could prove challenging for similar applications of other machine learning techniques. It is difficult to ensure that the toughest  
backgrounds are sufficiently well sampled in simulation if one has not already characterized the 
backgrounds which are easy or difficult to discriminate. It would be interesting to explore the importance of sufficient background sampling in the training of deep learning algorithms such as neural networks.

Our analysis has in a sense been optimized for the intermediate mass gap range, as we have trained the BDT with a single BSM model with 
mass splitting of $30 \gev$.  Unsurprisingly, a BDT trained on only this BSM model does far worse at identifying the presence of new physics 
if the mass splitting is either significantly larger or smaller, because of the changes to the underlying features of the kinematic distributions which drive 
this sensitivity  (see, for example~\cite{Dutta:2017nqv}).  It would be interesting to train a BDT with a range of mass splittings, to 
determine the trade-off between efficiency and generality of the analysis.  
In connection with this question, it would also be interesting to consider alternative choices in the training procedure.  For example, one could 
consider training BDTs separately to distinguish signal from each of the leading backgrounds, with the individual scores combined to yield an overall 
signal classifier.  In this work, we used a signal classifier threshold to make a binary classification of an event as either signal or background, but it 
would be interesting to consider alternative approaches.

{\bf Acknowledgements}  We are grateful to Alexander Khanov, Xerxes Tata and Evelyn Thomson for useful discussions.
We thank Jordan Pittman, Stephanie Samperio, Conrad Albrecht, and Joseph Hansen for technical assistance.
For facilitating portions of this research, B.~Dutta, J.~Kumar and J.W.~Walker wish to acknowledge the Center for Theoretical Underground Physics and Related Areas (CETUP*), The Institute for Underground Science at Sanford Underground Research Facility (SURF), and the South Dakota Science and Technology Authority for hospitality and financial support, as well as for providing a stimulating environment.
B.~Dutta is supported in part by  DOE Grant  DE-SC0010813.
T.~Ghosh  is supported by the funding available from the Department of Atomic Energy (DAE), Government of India for Harish-Chandra Research Institute (HRI).
J.~Kumar would like to thank the organizers of SUSY 2023 for their hospitality. 
J.~Kumar is supported in part by DOE grant DE-SC0010504.
P.~Sandick is supported in part by NSF grant PHY-2014075.
P.~Stengel would like to thank the organizers of Cosmology 2023 in Miramare for their hospitality. 
P.~Stengel is funded by the Instituto Nazionale di Fisica Nucleare (INFN) through the project of the InDark INFN Special Initiative: ``Neutrinos and other light relics in view of future cosmological observations'' (n. 23590/2021). J.W.~Walker is supported in part by the National Science Foundation under Grant NSF PHY-2112799.

\appendix
\section{Additional Validation Folds} \label{app:valid}
Plots representing additional validation folds with different selections of the events used for training and evaluation are provided here.  Fig.~\ref{fig:score_valid} shows the residual signal and background cross-section as a function of the classification score threshold.  Fig.~\ref{fig:discrimination_valid} shows the probability density distribution of the classification score for signal and background samples.  Fig.~\ref{fig:significance_valid} shows the evolution of the signal event count, signal to background ratio, and signal significance as a function of the classification score threshold.

\begin{figure}[htpb]
\centering
\includegraphics[width=0.475\linewidth]{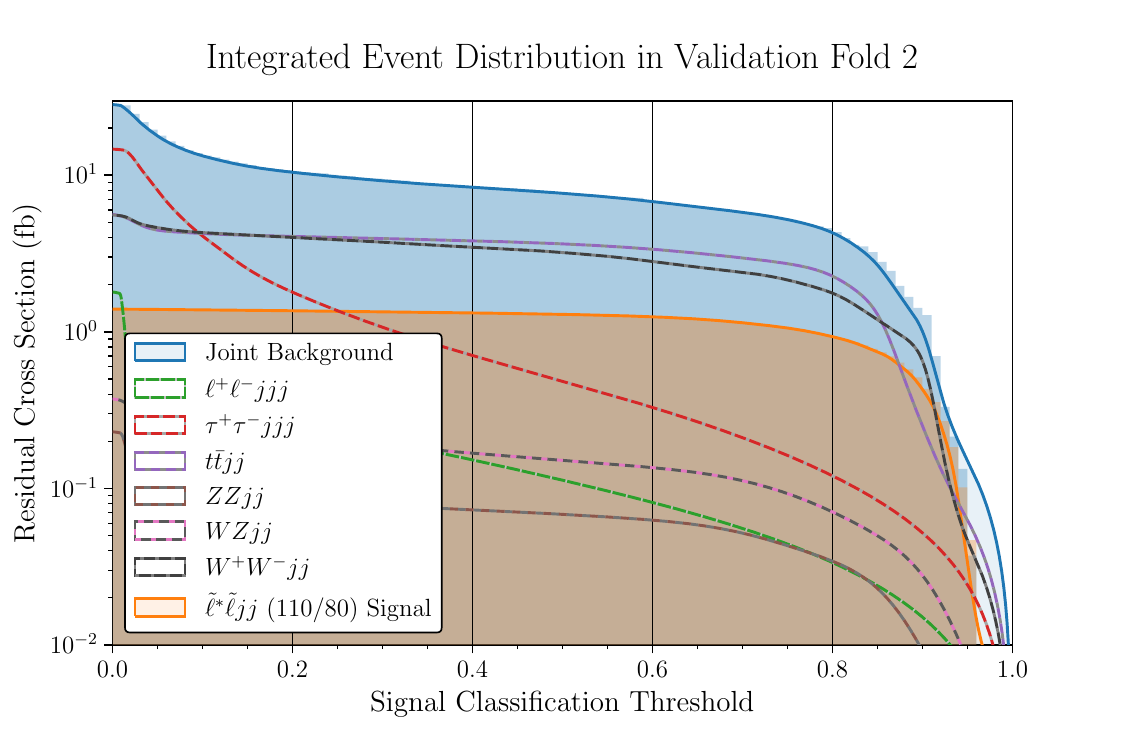}
\includegraphics[width=0.475\linewidth]{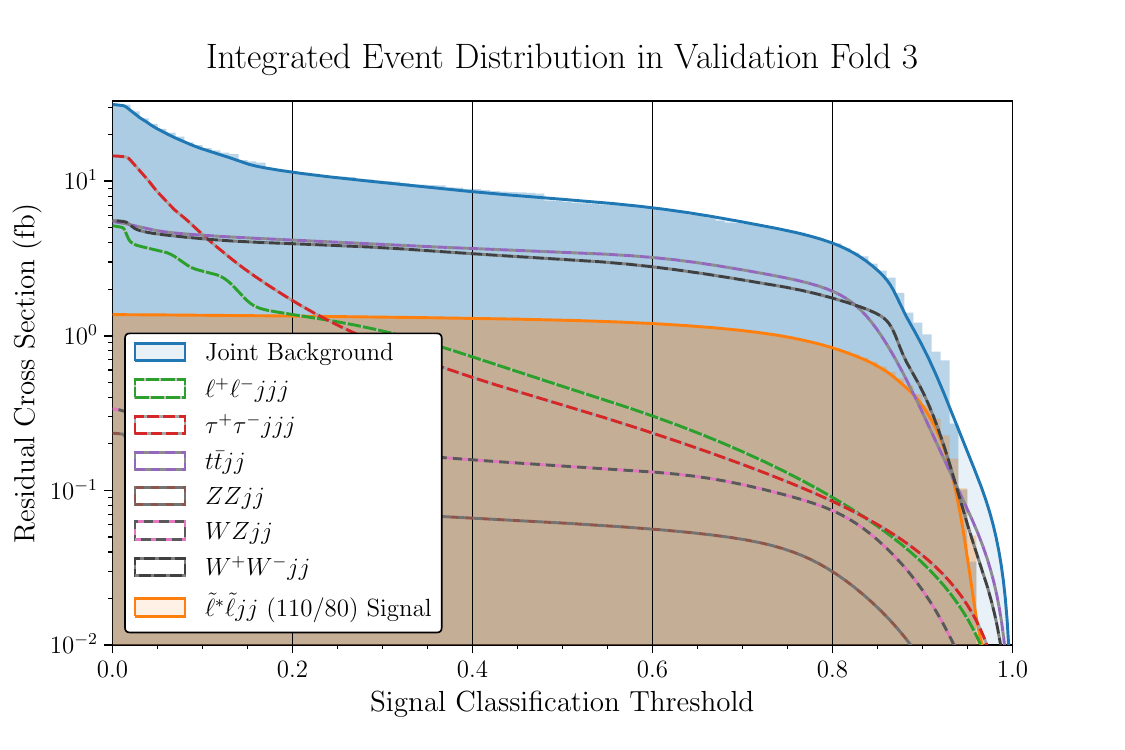} 
\caption{Same as Figure~\ref{fig:score}, but for different choices of training and evaluation samples. Plots illustrating the residual cross section for signal and background (as labeled)
as a function of the BDT classification score after precuts.}
\label{fig:score_valid}
\end{figure}

\begin{figure}[htpb]
\centering
\includegraphics[width=0.475\linewidth]{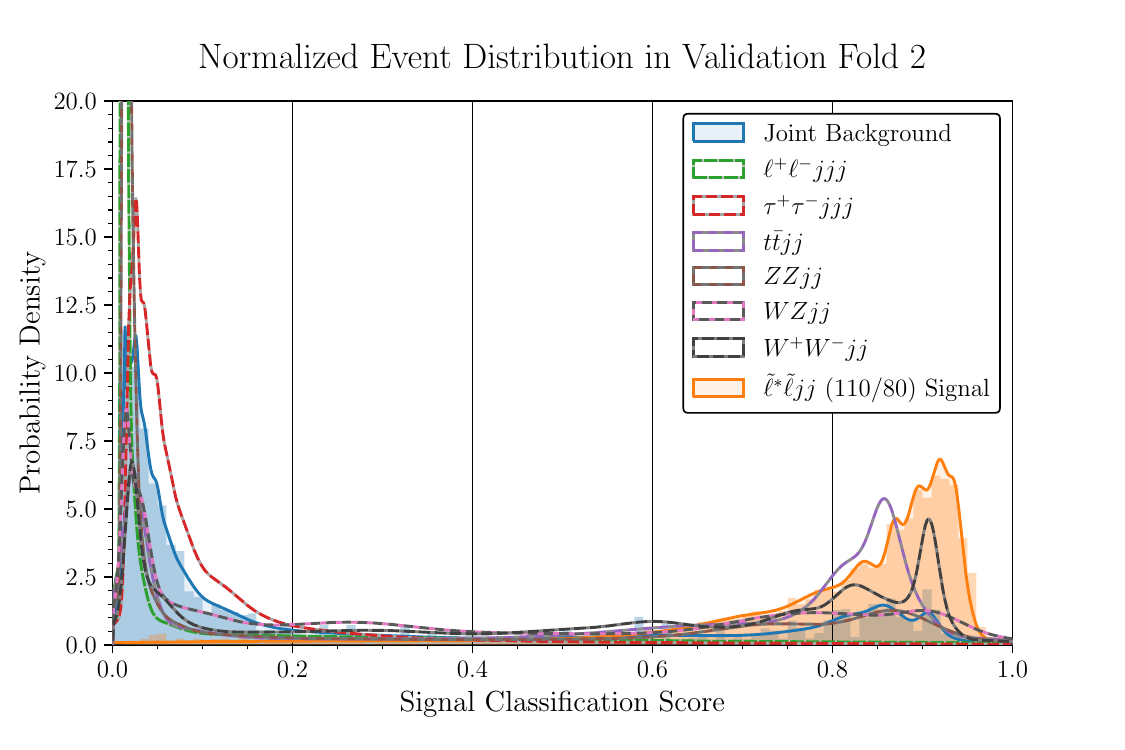} 
\includegraphics[width=0.475\linewidth]{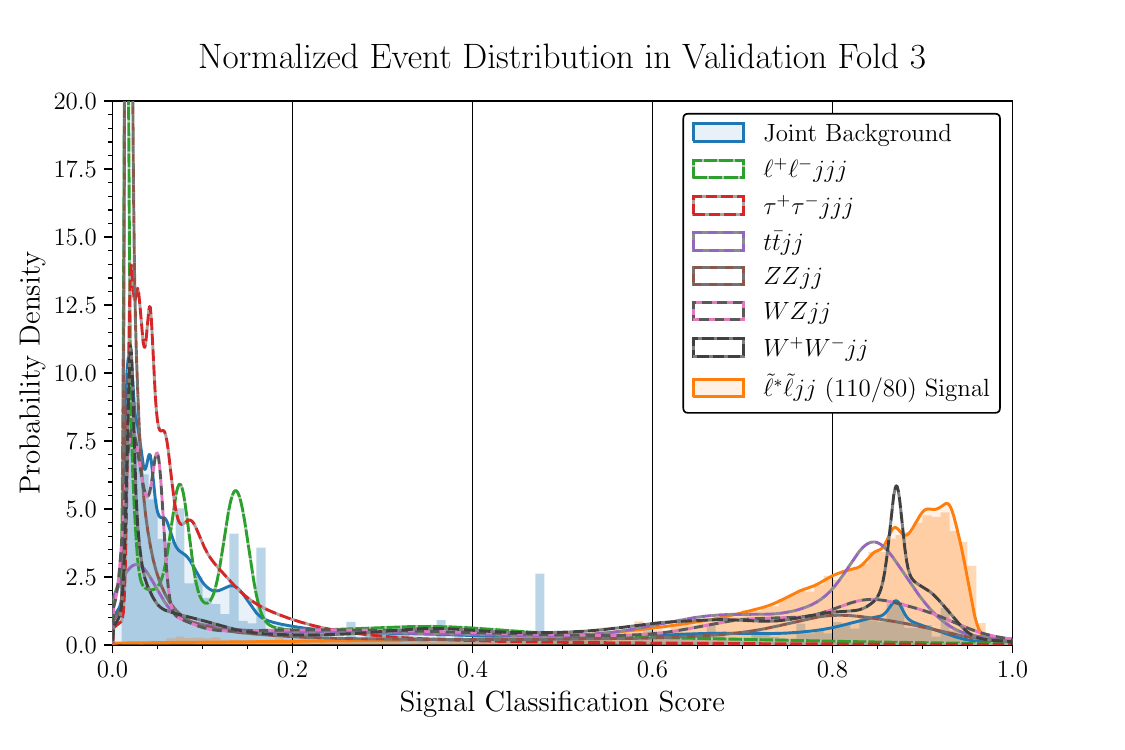} 
\caption{Same as Figure~\ref{fig:discrimination}, but for different choices of training and evaluation samples. Density plots representing the separation of signal from each background (as labeled) by
the boosted decision tree after precuts. For visual clarity, the densities are separately normalized such that the joint background density is not equivalent to the sum of the individual background densities.}
\label{fig:discrimination_valid}
\end{figure}

\begin{figure}[htpb]
\centering
\includegraphics[width=0.47\linewidth]{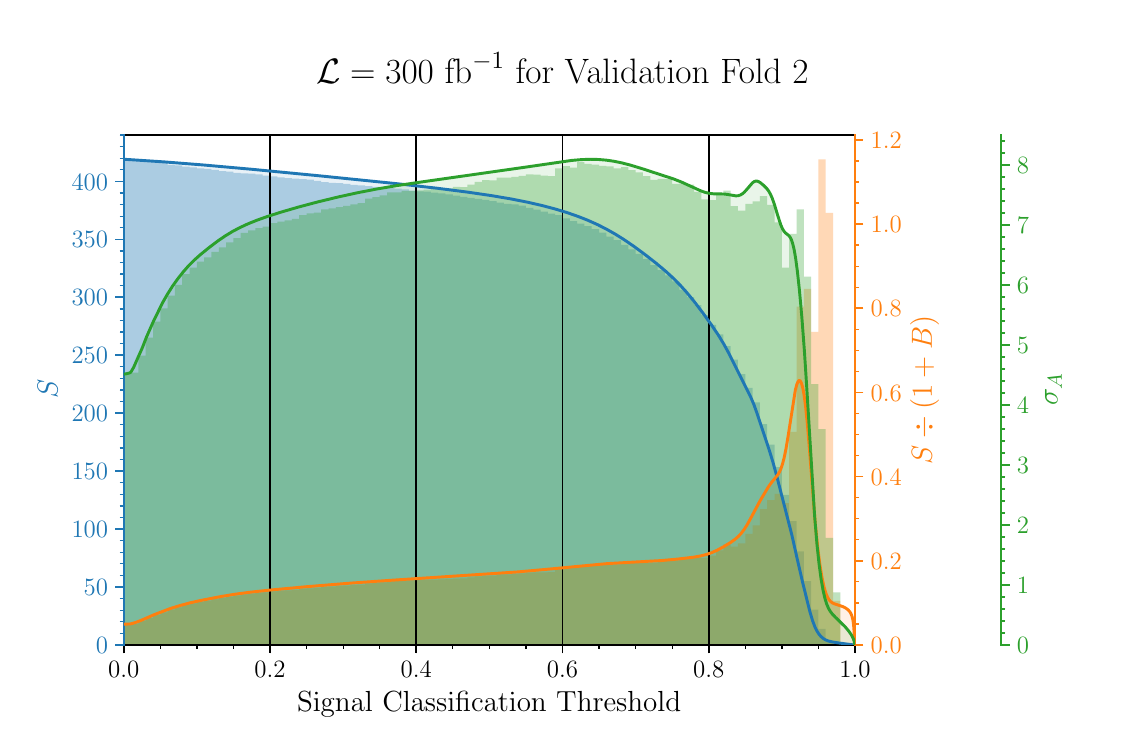} 
\includegraphics[width=0.47\linewidth]{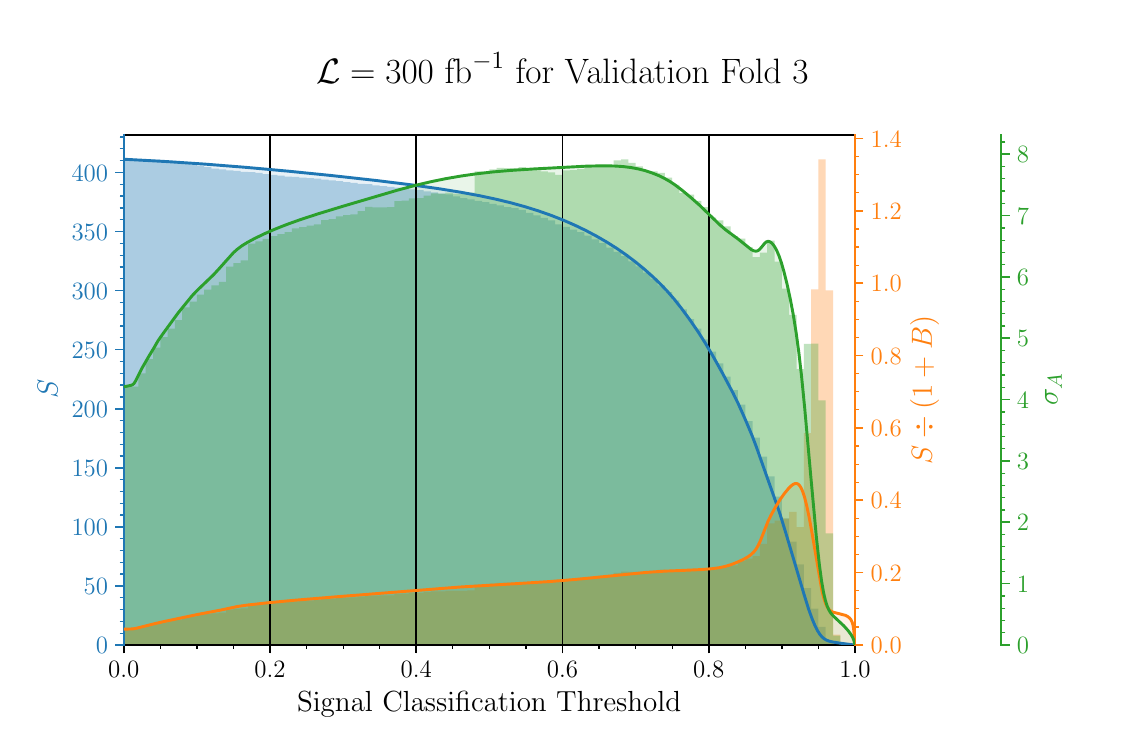} 
\caption{Same as Figure~\ref{fig:significance}, but for different choices of training and evaluation samples. The number of signal events ($S$, blue), signal-to-background ration 
($S\div(1+B)$, orange) and signal significance ($\sigma_A$, green) as a function of the 
signal classification threshold. Each curve represents an interpolation which smooths out the sharp 
features in the shaded region, which arise from small simulated event numbers.}
\label{fig:significance_valid}
\end{figure}

\section{$M_{\tau \tau}$}
\label{sec:Appendix_mtautau}

 In the process $pp \rightarrow Zj  \rightarrow j \bar \tau \tau$ 
the final state has two boosted $\tau$s with $(P^{\tau_1} + P^{\tau_2})^2 = M_Z^2$, and 
$\vec{P}_T^{\tau_1} + \vec{P}_T^{\tau_2} + \vec{P}_T^{j} =0$.  If each $\tau$ decays leptonically, with all products emitted in the approximately forward direction, then 
the observable particles will be two leptons, $\ell_{1,2}$, satisfying 
$\vec{P}_T^{\tau_i} = (1 +\xi_i) \vec{P}_T^{\ell_i}$, where $\xi_i$ is the ratio of momentum
carried by the pair of nearly collinear neutrinos to that of the visible lepton.
Since the $\vec{P}_T^{j, \ell_1, \ell_2}$ are observable, conservation of transverse momentum 
gives two equations for two unknowns:
$(1 +\xi_1) \vec{P}_T^{\ell_1} + (1 +\xi_2) \vec{P}_T^{\ell_2}+\vec{P}_T^{j} =0$.  One 
generally expects that one can solve for $\xi_{1,2}$, allowing one to reconstruct the 
mass of the parent $Z$-boson using 
$M_Z^2 = (1+\xi_1)(1+\xi_2)M_{\ell \ell}^2$, 
where $M_{\ell \ell}$ is the dilepton invariant mass.
Specifically, following the discussion in \cite{Dutta:2017nqv},
we demonstrate here that $M_{\tau\tau}$ from Eq.~\ref{eq:mtautau}
does indeed reconstruct the $Z$-boson mass under these circumstances.
Note that the conservation law implies 
$\vec{P}_T^{\tau_1} \times \vec{P}_T^{j}
= - \vec{P}_T^{\tau_2} \times \vec{P}_T^{j}
=- \vec{P}_T^{\tau_1} \times \vec{P}_T^{\tau_2}$.
From this, we have the following:

\begin{align}
M_{\tau \tau}^2
&= - M_{\ell \ell}^2 \frac{(\vec{P}_T^{\ell_1} \times \vec{P}_T^{j}) \cdot 
(\vec{P}_T^{\ell_2}\times \vec{P}_T^{j})}
{|\vec{P}_T^{\ell_1} \times \vec{P}_T^{\ell_2}|^2}
\nonumber\\
&= - M_{\ell \ell}^2 (1 +\xi_1)(1 + \xi_2) \frac{(\vec{P}_T^{\tau_1} \times \vec{P}_T^{j}) \cdot 
(\vec{P}_T^{\tau_2}\times \vec{P}_T^{j})}
{|\vec{P}_T^{\tau_1} \times \vec{P}_T^{\tau_2}|^2}
\nonumber\\
&=
 M_{\ell \ell}^2 (1 +\xi_1)(1 + \xi_2) \frac{(\vec{P}_T^{\tau_1} \times \vec{P}_T^{\tau_2}) \cdot 
(\vec{P}_T^{\tau_1}\times \vec{P}_T^{\tau_2})}
{|\vec{P}_T^{\tau_1} \times \vec{P}_T^{\tau_2}|^2}
\nonumber\\
&= M_{\ell \ell}^2(1+\xi_1)(1+\xi_2) = M_Z^2
\end{align}

\end{document}